\documentclass[letterpaper,11pt]{article}
\pdfoutput=1 

\usepackage{jheppub} 

\usepackage{tikz}
\usepackage{dsfont}
\usepackage{amsmath}
\def\ba{\begin{align}}\def\ea{\end{align}}
\def\beq{\begin{eqnarray}}\def\eeq{\end{eqnarray}}
\def\be{\begin{equation}}\def\ee{\end{equation}}
\def\ben{\begin{equation}}
\def\een{\end{equation}}
\def\bea{\begin{eqnarray}}
\def\eea{\end{eqnarray}}

\def\scry{{\cal I}}

\title{Superluminal Liouville walls in 2d String Theory and space-like singularities}

\author[1]{Sumit R. Das,}
\author[2]{Shaun D. Hampton,}
\author[3]{Sinong Liu.}

\affiliation[1]{Department of Physics and Astronomy, University of Kentucky, Lexington, KY 40506, U.S.A.}
\affiliation[2]{School of Physics,
Korea Institute for Advanced Study,
Seoul 02455, S. Korea}
\affiliation[3]{Yau Mathematical Sciences Center, Tsinghua University, Haidian District, Beijing 100084, China.}

\emailAdd{das@pa.uky.edu}\emailAdd{sdh2023@kias.re.kr}\emailAdd{sinongliu@mail.tsinghua.edu.cn}

\abstract{An interesting class of time dependent backgrounds in $1+1$ dimensional string theory involves worldsheet Liouville walls which move in (target space) time. When a parameter in such a background exceeds a certain critical value, the speed of the Liouville wall exceeds the speed of light, and there is no usual S-Matrix. We examine such backgrounds in the dual $c=1$ matrix model from the point of view of fluctuations of the collective field, and determine the nature of the emergent space-time perceived by these fluctuations. We show that so long as the corresponding Liouville wall remains time-like, the emergent space time is conformal to full Minkowski space with a time-like wall. However, for the cases where the Liouville wall is superluminal, the emergent space-time has a {\em space-like boundary} where the collective field couplings diverge. This appears as a space-like singularity in perturbative collective field theory. We comment on the necessity of incorporating finite $N$, as well as finite (double-scaled) coupling, effects to understand the behavior of the exact theory near this boundary.}

\begin{document}

\begin{flushright}
\end{flushright}

\maketitle
\flushbottom

\section{Introduction}
\label{sec:intro}

Two dimensional (bosonic) non-critical String Theory \cite{2d} is an early example of gauge-gravity duality which is understood to a significant extent on both sides. Here, the the internal degrees of freedom of gauged Matrix Quantum Mechanics of a single Hermitian matrix metamorphose into a space coordinate in the String Theory description.
This duality, originally established at the level of perturbative S-Matrices, has now been more recently tested at the non-perturbative level \cite{nonperturb, nonperturb2}.

It is also a rare instance in String Theory where one can treat certain time dependent backgrounds. The singlet sector of the dual double scaled matrix model is a theory of $N$ free fermions in an inverted harmonic oscillator potential, with the single particle Hamiltonian given by
\ben
h = \frac{1}{2} \left(p^2 - x^2\right)~.
\label{0-0}
\een
These fermions are (unstable) D-branes of the 2d String Theory \cite{newhat}.
In the large $N$ and classical limit, the phase space density $u(x,p,t)$ can be either zero or $1$, so that a state of the many fermion system is represented by a filled region in the single particle phase space. 
The ground state corresponds to the Fermi surface given by
\ben
x^2 - p^2 = 2\mu~,
\label{0-1}
\een
where ($-\mu$) is the scaled Fermi energy.

The time dependent backgrounds we will consider are represented by time dependent Fermi surfaces \cite{akk, alexandrov, strom, karc1, karc2, karc3, ddlm, dk,newtimedep, bck}. A widely studied class of examples which we will consider in this paper are fermi surfaces obtained by canonical deformations in the single particle phase space,
\ben
(x \pm p) \rightarrow (x \pm p) + \lambda_\pm e^{\pm rt} (x \mp p)^{r-1}~,
\label{0-2}
\een
where $r$ is a positive number. 
This leads to a fermi surface
\ben
x^2 - p^2 + \lambda_-e^{-rt} (x+p)^r + \lambda_+e^{rt}(x-p)^r + \lambda_+\lambda_- (x^2-p^2)^{r-1} = 2\mu~.
\label{0-3}
\een
For example, $r=1$ leads to a ``moving hyperbola'' Fermi surface \cite{strom}, while $r=2$ corresponds to a ``closing hyperbola'' or an ``opening hyperbola'' \cite{dk}. It is also possible to make general deformations which are linear combinations of the above for different values of $r$. 

In the world-sheet description, such backgrounds correspond to vertex operator deformations leading to a worldsheet potential \cite{akk,bck}
\ben
V_{WS} = - \mu \phi e^{2\phi} + \frac{\Gamma(r)}{\Gamma(1-r)} \int d^2\sigma e^{(2-r)\phi} \left(\lambda_+ e^{rt} + \lambda_- e^{-rt} \right)~.
\label{0-4}
\een
Here $\phi$ denotes the Liouville mode and $t$ the target space time. Backgrounds with $0 < r < 2$ contain an asymptotic region $\phi \rightarrow \infty$ and one may try to define an S-Matrix representing excitations coming in from this asymptotic region. In \cite{bck} this S-Matrix has been calculated from the worldsheet theory for $0 < r < 1$ and with $\lambda_- =0, \lambda_+ < 0$  when the Liouville wall receeds towards this asymptotic region at subluminal speed. The resulting S-Matrix displays, among other things, particle production. For $ 1 < r < 2$, the Liouville wall becomes superluminal after some time, and there is no asymptotic future. In this case it is less clear how to define observables and compute them from the worldsheet theory.

The main purpose of this paper is to throw light on the target space meaning of such time dependent solutions, particularly those with super-luminal Liouville walls, using collective field theory. 

In the matrix model, physical excitations are fluctuations of the density of eigenvalues, or the collective field $\partial_x \varphi$ and its canonical momentum $\Pi_\varphi$ \cite{jevsak}. In terms of the phase space density of fermions these are
\ben
\partial_x \varphi (x,t) = \frac{1}{2\pi} \int dp~ u(x,p,t)~,~~~~~~~
\Pi_\varphi \partial_x \varphi = -\frac{1}{2\pi g_s^2} \int  dp~p~u(x,p,t)~.
\label{0-5}
\een
Time dependent Fermi surfaces are  time dependent saddle points of the collective field. 

At the classical level, $u(x,p,t)$ is either one or zero: the boundary of a filled region is the Fermi surface. When a constant $x$ line intersects the Fermi surface at most twice (quadratic profiles), the collective field and its canonical momentum are determined by these intersection points \cite{polclass}. When a constant $x$ line intersects the Fermi surface more than twice one needs to introduce additional fields in this classical bosonization, though at the full quantum level a single bosonic field may suffice. Such non-quadratic profiles represent states whose quantum fluctuations do not vanish in the classical limit we are considering \cite{dmat, ddbrane}.

The fluctuations around the ground state saddle describe a massless scalar in $1+1$ dimensions in \cite{dasjev}. One can always go to a coordinate system where the metric is conformal to Minkowski space-time: in these coordinates the global properties of the emergent space-time become transparent. For the ground state the Minkowskian space coordinate is the time of flight of a particle from a point $x$ to the potential, while the Minkowskian time is the time of the Matrix Model. This massless scalar is related to the single dynamical mode of the two dimensional string theory, the ``massless tachyon''. Their relationship involves a non-local transform, the non-locality being at the string scale \cite{polnat}. In this paper we will not be concerned with non-locality at the string scale.

As reviewed below, this passage to conformally Minkowski coordinates continues to hold for fluctuations around a time dependent background.  Once expressed in these coordinates, the time dependence appears in the coupling. The exactly solvable cases $r=1$ and $r=2$ represent quadratic profiles and two intersection points at  $p=P_\pm (x,t)$ can be determined by solving \eqref{0-3}, so that $\partial_x \varphi$ and $\Pi_\varphi$ can be determined in terms of $P_\pm (x,t)$. The physics of fluctuations can be then determined. 
One finds that for $r=1$ there is particle production from a moving mirror \cite{strom, karc1, karc2,karc3,ddlm}. For $r=2$, however, one gets emergent space-times with space-like boundaries at late Matrix Model times, where the couplings become large \cite{dk}. From the point of view of (perturbative) collective field theory this would appear as a space-like singularity.

Similar space-times also result from a quantum quench of the matrix model, obtained by a time dependent coupling which is equivalent to a time dependent coefficient of the inverted oscillator potential \cite{dhl}. There are other classes of time dependent solutions \cite{takayanagi} in the $c=1$ model which represent the duals of 2d string theory with a timelike dilaton matter \cite{stromtaka} on the world sheet.

For general non-integer $r$, the algebraic equation \eqref{0-3} cannot be solved analytically, and the above strategy of deriving the collective field theory becomes problematic. In this paper we will use a combination of analytical reasoning and numerical calculations to determine some key causal features of the space-time which emerges from these time dependent solutions for general $r$. We will deal with $\lambda_- =0$ so that we start in the ground state at early times. The Fermi surface is given by
\ben
x_+x_-  + \lambda_+ e^{rt} x_-^r = 2\mu~,
\label{1-1}
\een
where we have used the phase space variables
\ben
x_\pm = x \pm p~.
\label{1-2}
\een
We prove that when $\lambda_+ < 0$ the Fermi surface {\em always} has a quadratic profile, so that the classical derivation of the collective field is standard \cite{polclass}.
We find that for $ 0 < r < 1$ the space-time is geodesically complete, and the couplings are strong along a time-like line. On the other hand, for $ r > 1$ we show that there is a space-like boundary with strong coupling: similar to a space-like singularity from the point of view of collective field excitations. 

Of course this is a singularity in {\em perturbative} collective field theory. What this means is that the space-time interpretation of the model, which is directly connected to worldsheet string theory, needs modification from strong coupling effects. This does not necessarily mean that there is something pathological in the fermion theory itself. Nevertheless, there is something pathological in the fermionic theory in a double scaled potential which is unbounded. The double scaling limit is performed by starting with fermions in a potential which is bounded from below at both sides and taking $N \rightarrow \infty$ keeping a scaled fermi level finite. In this limit the IR walls appear at $O(\sqrt{N})$ away from the potential, which gets pushed to infinity. As we will see below, in the time dependent solutions we are considering the fermions gets pushed to the region of this IR wall at late times and the reflection from the wall becomes appreciable: this is the time when the collective field couplings become large.  This means that we need to consider finite $N$ effects to understand the true time evolution at such late times. We therefore need to understand not only strong string coupling effects, but finite $N$ effects as well. (In these models these are two different effects.)

For $ \lambda_+ > 0$ and $r > 2$ we find non-quadratic profiles and fermions spill over from one side of the potential to the other. This means that we need to consider the Fermi surfaces on both sides. However we find that at late times a constant $x$ line intersects each Fermi surface once, and a collective field can be defined at the classical level following \cite{dk,dhl}. We defer a treatment of this case to future work. We expect the results to be qualitatively similar to the exactly solvable cases.

In Section \ref{exact-solvable} we develop the strategy of determining the emergent space-time perceived by collective field fluctuations and review some exactly solvable cases of $r=1,2$ which have been discussed in earlier literature. Section \ref{fermi-sf-general} is devoted to a discussion of the nature of the fermi surfaces for arbitrary $r$. Section \ref{emergent-spt} derives the Minkowskian coordinates and global properties of the emergent space-time and shows that there are space-like boundaries for $r > 1$. In section \ref{divcp} we compute the cubic couplings on these space-like boundaries and show that they diverge. Section \ref{discuss} discusses the issues which need  to be understood to really understand what is going on here. Details of the derivation of various formulae in the main text are given in the appendices.

In this paper we use units where $2\mu = 1$.

\section{The strategy and exactly solvable cases}\label{exact-solvable}

In this section we review the general strategy of determining the emergent space-time structure perceived by fluctuations around a Fermi surface \eqref{0-3}, and review the results for $r = 1,2$. We will work in the classical limit.

For $\lambda_- =0$ the Fermi surface can be parametrized by a parameter $\alpha$,
\bea
x & = & \cosh \alpha - \frac{\lambda_+}{2}e^{rt} e^{-(r-1)\alpha}~, \nonumber \\
p & = & \sinh \alpha - \frac{\lambda_+}{2}e^{rt} e^{-(r-1)\alpha}~.
\label{2-1}
\eea
For quadratic profiles there are at most two solutions for $\alpha$ for a given $x$ which we denote by $\alpha_\pm$ so that there are two intersections $P_\pm (x,t) = p(\alpha_\pm, t)$ for a given $x$. The collective field and the conjugate momenta are then determined by 
\ben
\partial_x \varphi = \frac{1}{2\pi} \left( P_+(x,t) - P_-(x,t) \right)~,\qquad
\Pi_\varphi \partial_x \varphi = - \frac{1}{4\pi g_s^2} \left( P_+(x,t)^2 - P_-(x,t)^2 \right)
\label{1-8}
\een
with the classical Hamiltonian given by
\ben
H = \int dx ~\left[ \frac{g_s^2}{2} \Pi_\varphi (\partial_x \varphi) \Pi_\varphi + \frac{\pi^2}{6 g_s^2} (\partial_x \varphi)^3 - \frac{x^2-1}{2 g_s^2} \partial_x \varphi \right]~,
\label{1-9}
\een
leading to an action
\ben
S = \frac{1}{g_s^2} \int dt dx~\left[ \frac{(\partial_t \varphi)^2}{\partial_x \varphi} - \frac{\pi^2}{6} (\partial_x \varphi)^3 + \frac{x^2-1}{2} (\partial_x \varphi) \right]~.
\label{1-10}
\een
Using Hamiltonian equations of motion we can now re-express the relations \eqref{1-8} as
\ben
\partial_x \varphi = \frac{1}{2\pi }\left[ P_+(x,t)-P_-(x,t) \right]~, \qquad
\frac{\partial_t \varphi}{\partial_x \varphi}= - \frac{1}{2}
\left[ P_+(x,t) + P_-(x,t) \right] ~.
\label{1-8a}
\een
The time dependent Fermi surfaces correspond to time dependent classical solutions $\partial_x \varphi_0 (x,t)$. Fluctuations around the classical solution are
\ben
\varphi (x,t) = \varphi_0 (x,t) + \frac{g_s}{\sqrt{\pi}} \eta (x,t) ~.
\label{1-11}
\een
The action for $\eta (x,t)$ is non-polynomial. The quadratic and cubic pieces are
\ben
S^{(2)} = \frac{1}{2\pi} \int dx dt \left[ \frac{(\partial_t \eta)^2}{\partial_x\varphi_0}
-2 \frac{(\partial_t \varphi_0)}{(\partial_x \varphi_0^2)} (\partial_t \eta)(\partial_x\eta) + \left( \frac{(\partial_t \varphi_0)^2}{(\partial_x \varphi_0)^3} - \pi^2 \partial_x\varphi_0 \right) (\partial_x \eta)^2 \right] ~,
\label{2-11}
\een
while the cubic interaction part is 
\ben
S^{(3)} = -\frac{1}{\pi^{3/2}}\int dx dt \left[ \frac{1}{2} \frac{1}{(\partial_x \varphi_0)^2} (\partial_t \eta)^2 (\partial_x \eta)- \frac{\partial_t \varphi_0}{(\partial_x \varphi_0)^3}(\partial_t \eta)(\partial_x \eta)^2 + \left( \frac{(\partial_t \varphi_0)^2}{(\partial_x \varphi_0)^4} + \frac{\pi^2}{6} \right) (\partial_x \eta)^3 \right] ~.
\label{2-12}
\een
The quadratic action shows that the fluctuation field is a massless field which is propagating on a $1+1$ dimensional space-time with a metric which is conformal to 
\ben
ds^2  =  -dt^2 + \frac{(dx + \frac{\partial_{t} \varphi_0}{\partial_x \varphi_0}d t)^2}{(\pi \partial_x \varphi_0)^2} ~.
\label{2-13}
\een
Since we are dealing with a massless field in $1+1$ dimensions the quadratic action is insensitive to a conformal transformation of the metric. The signature of the metric \eqref{2-13} remains negative at all times,
\ben
{\rm det}(g) = -\frac{1}{\pi^2(\partial_x \varphi_0)^2}~.
\een
To look at the global properties of the emergent space-time it is convenient to go to coordinates $(q,\tau)$ in which the metric is conformal to Minkowski space-time. For quadratic profiles, \cite{alexandrov} has shown that these coordinates are simply given in terms of the two solutions $(\alpha_+, \alpha_-)$ of \eqref{2-1} which lead to the same value of $x$
\ben
\tau = t - \frac{1}{2} (\alpha_+ + \alpha_-)~, \qquad q = \frac{1}{2} (\alpha_+ - \alpha_-)~.
\label{2-14}
\een
When $\lambda_\pm = 0$, symmetry dictates that $\alpha_+ = - \alpha_-$ leading to $\tau = t$ and $q = \cosh^{-1} x$ is the ``time of flight'' coordinate \cite{dasjev}. The couplings are strong at $q=0$ at all times.

The general strategy is then the following. Given a quadratic time dependent Fermi surface we first determine the $\alpha_{\pm}$, and then determine $(\tau,q)$ by \eqref{2-14}. This transformation holds the key to the global features of the emergent space-time.

\subsection{$r=1$: Moving mirrors}

The first case where this strategy can be implemented analytically is $r=1$ \cite{strom, karc1, karc2}. The Fermi surface is a hyperbola whose focus is shifted by a time dependent amount compared to the ground state. It is easy to check that the two solutions of \eqref{2-1} for a given $x$ are
\ben
\alpha_+ = - \alpha_- = \log \left[ \left(x + \frac{\lambda_+}{2}e^t \right) + 
\sqrt{\left(x + \frac{\lambda_+}{2}e^t \right)^2 -1} \right]~,
\label{2-15}
\een
so that the Minkowskian coordinates are given by
\ben
q = \cosh^{-1} \left( x + \frac{\lambda_+}{2}e^t \right)~, \qquad \tau = t~.
\label{2-16}
\een
The Penrose diagram of the emergent space-time is exactly the same as in the vacuum. The two sides of the potential are separated by a solid barrier which is represented in these coordinates by $q=0$. Excitations come in from $\scry^-$, get reflected back to $\scry^+$ from this mirror. However in the original coordinates the mirror is moving, which leads to particle production \cite{ddlm}.

\subsection{$r=2$: Space-like singularities}

The emergent space-time is non-trivial when $r=2$ \cite{dk}. 
In this case the solutions for $\alpha_\pm$ are
\ben
e^{\alpha_\pm} = x \pm \sqrt{x^2 - (1-\lambda_+ e^{2t})}~.
\label{2-17}
\een
The Minkowskian coordinates defined in \eqref{2-14} then become
\ben
\tau =  t -\frac{1}{2} \log \left( 1 - \lambda_+ e^{2t}\right)~, \qquad
\cosh q =  \frac{x}{\sqrt{1- \lambda_+ e^{2t}}}~.
\label{2-18}
\een
The two intersections of the Fermi surface with a constant $x$ line, $P_\pm(x,t)$ are then given by
\ben
P_\pm (x,t)  = \sinh \alpha_\pm -\frac{\lambda_+}{2} e^{2t-\alpha_\pm}
 =  x - \frac{1}{x \pm \sqrt{x^2-(1-\lambda_+ e^{2t})}}~.
\label{2-19}
\een
Using \eqref{1-8} and \eqref{2-18} we can read off the collective field and the canonically conjugate momentum,
\ben
\partial_x \varphi_0  =  \frac{\sinh q}{\pi\sqrt{1-\lambda_+e^{2t}}}~, \qquad
g_s^2 \Pi_\varphi  =  \frac{\partial_t \varphi_0}{\partial_x \varphi_0} = \frac{(\cosh q) \lambda_+ e^{2t}}{\sqrt{1-\lambda_+e^{2t}}}~.
\label{2-20}
\een
The space-time structure depends on the sign of $\lambda_+$.

\subsubsection{$\lambda_+ < 0$}

When $\lambda_+ < 0$, it will be convenient to shift the origin of time $t \rightarrow t- \frac{1}{2} \log (|\lambda_+|)$. This sets $\lambda_+ = -1$. The equation \eqref{2-18} shows that a constant $t$ surface is also a constant $\tau$ surface. However at the end of time evolution in the Matrix Model, $t \rightarrow \infty$, the Minkowskian $\tau$ approaches a {\em finite} value $\tau \rightarrow 0$. The corresponding Penrose diagram ends at $\tau = 0$. This would be a geodesically incomplete space-time and normally one would continue beyond this  value. However the fundamental description of the theory is the Matrix Model, and there is no meaning of continuing the time evolution beyond $t = \infty$.

We therefore conclude that the fluctuations perceive a space-time with a space-like boundary. 
The cubic couplings of the field $\eta$ diverge all along this line. This may be seen from \eqref{2-20} and the couplings which appear in \eqref{2-12}. For $t \rightarrow \infty$ and for any $q$
\ben
\frac{1}{(\partial_x \varphi_0)^2} \sim \frac{1}{(\sinh q)^2 (1-e^{2\tau})}~, \quad
\frac{(\partial_t \varphi_0)}{(\partial_x \varphi_0)^3} \sim \frac{\cosh q~ e^{2\tau}}{(\sinh q)^2 (1-e^{2\tau})^{3/2}}~, \quad
\frac{(\partial_t \varphi_0)^2}{(\partial_x \varphi_0)^4} \sim \frac{e^{4\tau}}{(\tanh q)^2(1-e^{2\tau})^2}~.
\label{2-21}
\een
The third cubic coupling diverges all along the line $\tau =0$, while the others diverge along portions of this line. 
From the point of view of the classical collective theory this is similar to a space-like singularity. This is of course the end point of time evolution in the matrix model. In a later section we will discuss what aspects of the matrix model come in to play in understanding this region.

For future considerations it is useful to derive these results by first going to the late time limit. Let us define a coordinate $y$
\ben
x = \cosh (y)~.
\label{2-22}
\een
We will look at the asymptotic region
\ben
y \gg t \gg 1~.
\label{2-23}
\een
In this limit we can expand the square-root in \eqref{2-17} as 
\ben
\sqrt{x^2 -(1+ e^{2t})} \sim x - \frac{e^{2t}}{2x} = \frac{1}{2}  e^y - e^{2t-y} ~.
\label{2-24}
\een
Using this in (\ref{2-17}) we get, in this asymptotic region,
\ben
\alpha_+ \approx y~,\qquad \alpha_- \approx 2t - y~.
\label{2-25}
\een
The Minkowskian coordinate $q$ is then given by
\ben
q \approx y-t~.
\label{2-27}
\een
This agrees with the limiting form of the exact expression \eqref{2-18}. In the expression for $P_\pm (x,t)$ we use \eqref{2-25},
in the regime \eqref{2-23},
\bea
P_+ & \approx & \frac{1}{2} \left( 2\sinh (y)+ e^{2t-y} \right) = \frac{1}{2} \left( 2\sinh(q+t) + e^{t-q} \right)~, \nonumber \\
P_- & \approx & \frac{1}{2} \left( -2 \sinh(y-2t) + e^y \right) = \frac{1}{2} \left( -2\sinh(q-t) + e^{q+t} \right)~.
\label{2-28}
\eea
This leads to (for $t \gg 1$)
\ben
\partial_x \varphi_0 = \frac{1}{2\pi} (\sinh q)  e^{-t}~, \qquad
\frac{\partial_t \varphi_0}{\partial_x \varphi_0} = -\frac{1}{2}(\cosh q) (2e^t-e^{-t}) \sim (\cosh q) e^{t}~.
\label{2-29}
\een

\subsubsection{$\lambda_+ > 0$}

The nature of the space-time is rather different when $\lambda_+ > 0$. 
Once again, we shift the origin of time to set $\lambda_+=1$. Now there is a finite value of the time $t = 0$ at which $\tau = +\infty$. At this time, the mirror at $q=0$ disappears and excitations cross over from one side of the potential to the other. Note that this is a null line where the couplings are finite: from the point of view of the emergent space-time this is $\scry^+$. Normally this would be the end of space-time. However now the Matrix Model instructs us to continue the time evolution beyond  this time. For $t > 0$, one needs to attach another piece of space-time which does not have a barrier at $q=0$. The way this is done is discussed in detail in \cite{dk,dhl}.

The above picture emerges strictly in the double scaled limit where the number of fermions, $N$, is infinite. The modification of this picture for finite $N$ will be discussed in a later section.

\section{Fermi surface profiles for general $r$} \label{fermi-sf-general}

For generic $r$, we encounter two issues. First, as we will see below, the profile of the Fermi surface may not be quadratic so that a constant $x$ line intersects it more than twice. In this case, we need to figure out how to determine the collective field theory at the classical level. Secondly, even for the cases where the profile is quadratic, the equations (\ref{2-1}) cannot be solved analytically: we need to develop suitable approximation methods in physically interesting regimes. We will discuss the first issue in a later section, and deal with the second issue in this section. 

We will first prove that the profile is quadratic for $\lambda_- =0$ and $\lambda_+ < 0$. Then we will proceed to determine the two intersections of a constant $x$ line in the late time limit $t \gg 1$.

\subsection{Quadratic profiles for any $r$ with $\lambda_-=0$ and $\lambda_+ < 0$}

In this subsection we will show that when $\lambda_-=0$ and $\lambda_+ < 0$, the Fermi surface has a quadratic profile for all values of $r$.

To simplify the algebra, we will set $\lambda_+ = -1$, by shifting the origin of time. The original time variable can of course be easily restored. 

This equation \eqref{1-1} makes sense for the region $x_- > 0$. Furthermore \eqref{1-1} and \eqref{1-2} lead to 
\ben
2x = \frac{1}{x_-} + x_- +e^{rt} x_-^{r-1} ~.
\label{1-3}
\een
However for $x_- > 0$ we have
\ben
\frac{1}{x_-} + x_-  \ge 2~.
\label{1-4}
\een 
So that \eqref{1-3} now implies $x > 1$. We can therefore parametrize
\ben
x_- = e^\theta~, \qquad x = \cosh \psi~.
\label{1-5}
\een
The equation \eqref{1-3} then leads to 
\ben
rt - \log (2) + (r-1)\theta = \log (\cosh \psi - \cosh \theta)~.
\label{1-6}
\een
Note that \eqref{1-3} requires 
\ben
\cosh \psi > \cosh \theta ~.
\label{1-7}
\een
Consider the function $F(\theta, \psi) = \log (\cosh \psi - \cosh \theta)$. Clearly $F(\theta,\psi) = F(-\theta, \psi)$. It is straightforward to check that for a given $\psi$, the function $F(\theta, \psi)$ as a function of $\theta$ has a single maximum at $\theta = 0$ and monotonically decreases with increasing $|\theta |$ on both sides. The left hand side of \eqref{1-6} is, however linear in $\theta$. This means that for any given $t,\psi$ there are at most two solutions for $\theta$, i.e. two solutions for $x_-$, obeying \eqref{1-6}. This in turn shows that  a constant $x$ line intersects the Fermi surface twice, except at one point where it is tangential to the surface.

Comparing \eqref{1-3} with \eqref{2-1} it is clear that
\ben
x_-^\pm = e^{\theta_\pm} = e^{-\alpha_\mp}~.
\label{4-0}
\een

Since the fermi surface does not cross over from $x > 0$ to $x < 0$ we will restrict our attention to the region $ x > 0$. The fermi surfaces for the region $x < 0$ can be obtained by simple reflections about the $x$ and the $p$ axes.

\begin{figure}[!h]
\centering
\includegraphics[width=2.5in]{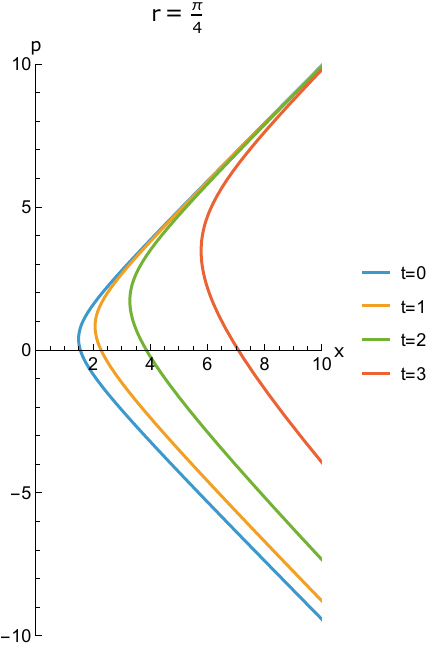}
\caption{Fermi surfaces at various times for $r = \frac{\pi}{4}$.}
\centering
\label{rlessone}
\end{figure}

The nature of the Fermi surface depends on the value of $r$. For $r < 1$ the Fermi surface roughly retains its shape, with its edge moving towards the large $x$ asymptotic region.
Figure \ref{rlessone} shows the profile of the fermi surface at various times for $\lambda_+ = -1$ and $r=\frac{\pi}{4}$. As time proceeds, the edge moves to the right, and to positive values of $p$. The upper branches coincide at large values of $x$. The lower branch always intersects the $x$ axis, with the point of intersection moving to the right as time proceeds. This is qualitatively similar to the exactly solvable case $r=1$ reviewed above. This is typical of any $0 < r < 1$.

\begin{figure}[!h]
\centering
\includegraphics[width=2.5in]{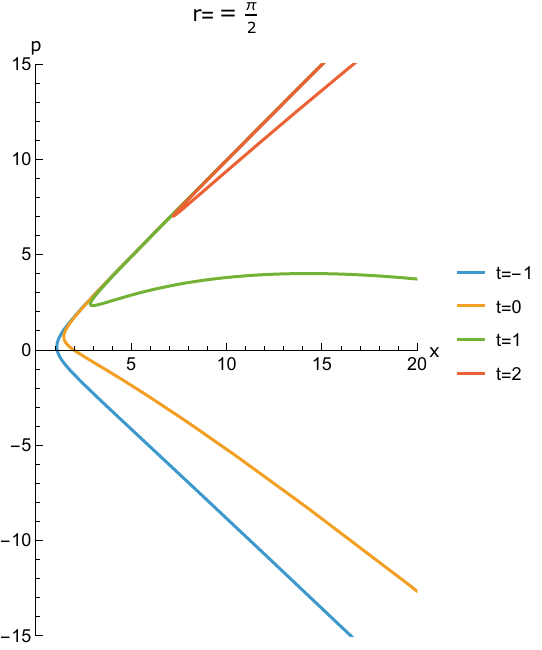}
\caption{Fermi surfaces at various times for $r = \frac{\pi}{2}$.}
\centering
\label{ronetwo}
\end{figure}

For $1 < r < 2$ the upper and lower branches come close together at late times.  The lower branch is now a concave line, but eventually intersects the $x$ axis, at larger and larger values of $x$ as time proceeds. 
Figure \ref{ronetwo} shows the profile of the fermi surfaces at various times for $r = \frac{\pi}{2}$. The general behavior is similar for any $r$ in this range.  

\begin{figure}[!h]
\centering
\includegraphics[width=2.5in]{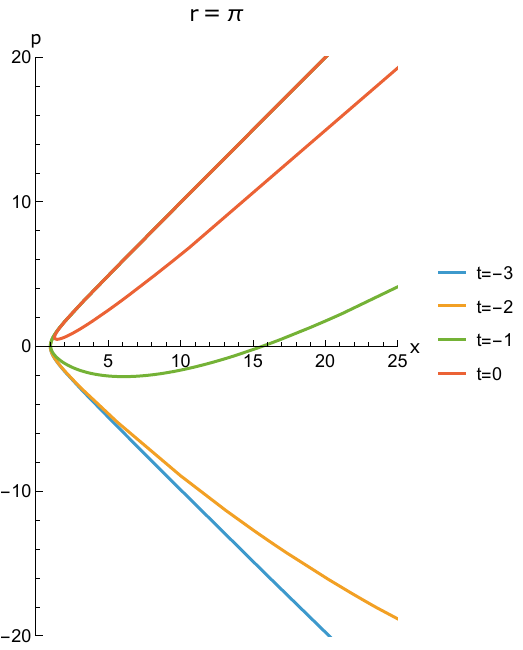}
\caption{Fermi surfaces at various times for $r = \pi$.}
\centering
\label{rgtwo}
\end{figure}

For $r > 2$ the lower branch is convex and at sufficiently late times never intersects the $x$-axis. This is shown in Figure \ref{rgtwo} for $r = \pi$.

In the next subsection we will obtain the explicit solution for $x^-(\psi)$ as a power series and obtain useful approximate solutions at late times.

\subsection{Approximate solutions in various regimes}

The equation \eqref{1-3} can be re-written, using the definitions \eqref{1-5}
\ben
(x_- - e^{\psi})(x_- - e^{-\psi}) + e^{rt}x_-^r = 0~.
\label{4-1}
\een
In the previous subsection we showed that this equation has at most two solutions. In Appendix \ref{app-a} we prove that these solutions can be exactly expressed as a power series,                               
\ben
x_-^\pm = e^{\pm \psi} \left[1 + \sum_{n=1}^\infty \frac{b_n^\pm}{n !} \left( \frac{ e^{ r ( t \pm  \psi )} }{1 -  e^{\pm 2 \psi} } \right)^n  \right]~,
\label{4-2}
\een
where, by observation,
\ben
b_n^{\pm} =  \sum_{k=1}^{n} \frac{ (n +k -2 )! }{(k-1)! (n-k)!} (nr)_{n-k} \left[  \frac{ e^{\pm 2 \psi} }{1 -  e^{\pm 2 \psi} } \right]^{k-1} ~.
\label{4-3}
\een
We will now show how the series solution \eqref{4-2} can be used to obtain 
$\theta_\pm = -\alpha_\mp$ in the asymptotic region with large values of $x$ and $t$.

\subsubsection{Solution near the turning point}

The Fermi surface has a turning point at some value of the parameter $\alpha_\star$ where $\partial_\alpha x(\alpha,t) = 0$ or equivalently
\begin{equation}
\partial_- \left[ \frac{1}{2}\left( x_- + x_-^{-1}\right) + \frac{1}{2} e^{rt} x_-^{r-1} \right] \bigg|_{x_-=x_-^*}=0 ~.
\label{turnpt}
\end{equation}
This leads to the solution
\begin{equation}
x_-^* =
\begin{cases}
\exp \left( \frac{\log (1-r) + rt}{2-r}\right)= (1-r)^{\frac{1}{2-r}} e^{\frac{r}{2-r} t }~, & r<1 \\
1 ~, & r =1 \\
\exp \left(- \frac{\log (r-1) + rt}{r}\right) = (r-1)^{-\frac{1}{r}} e^{-t} ~, & r>1 \\ 
\end{cases}~.
\label{minimalF}
\end{equation}
Expanding around this point, $x = x_-^* + \delta x$ for small $\delta x$ we get
\begin{equation}
x_-^{\pm} = x_-^* \left[ 1 \pm \left( \frac{2\delta x}{ (x_-^*)^{-1}  +\frac{(r-1)(r-2)}{2} e^{rt} (x_-^*)^{r-1} } \right) + \mathcal{O} (\delta x) \right]~.
\label{solnnear}
\end{equation}
This leads to the following results for $\alpha_\pm$
\begin{eqnarray}
\alpha_+ &=& - \log x_-^{-}  \approx -\log x_-^* + \left( \frac{ 2\delta x }{ (x_-^*)^{-1} + \frac{(r-1)(r-2)}{2} e^{rt} (x_-^*)^{r-1}  } \right)^{1/2}~, \label{smallxw1} \\
\alpha_- &=&  - \log x_-^{+} \approx  -\log x_-^* - \left( \frac{ 2 \delta x}{  (x_-^*)^{-1} + \frac{(r-1)(r-2)}{2} e^{rt} (x_-^*)^{r-1}  } \right)^{1/2} \label{smallxw2}~.
\end{eqnarray}
Finally, in Appendix \ref{app-b} we prove that 
\ben
x(\alpha_\star) < \frac{1}{2} + e^t~.
\label{4-4}
\een

\subsubsection{ Solution far from the turning point} \label{solnfar}

We now turn to the solution at late times and for values of $x$ far from the turning point. Equation \eqref{4-4} implies that this region is characterized by
\ben
\psi \gg t \gg 1~.
\label{4-5}
\een
Consider now the series solution for $x_-^{-}$. From \eqref{4-2} we immediately conclude that for any $r > 0$ in the asymptotic region \eqref{4-5}
\ben
x_-^- \sim e^{-\psi} \implies \alpha_+= -\log (x_-^-) \sim \psi~.
\label{4-6}
\een
The expression for $\alpha_-$ depends on the values of $r$.

First consider the range $0 < r < 2$. 
For large $t,\psi$ the power series expansion of $x_-^+$ in \eqref{4-2} is in the quantity 
\ben
e^{(r-2)\psi + rt}~.
\label{4-7}
\een
This means that when $r < 2$ and
\ben
rt \ll (2-r)\psi~,
\label{4-8}
\een
this expansion is in powers of a small quantity. Therefore in the regime \eqref{4-8} the leading order result for $\log (x_-^+) = -\alpha_-$ is given by
\ben
\alpha_- = -\psi = -\cosh^{-1} (x)~, \qquad {\rm for}~~rt \ll (2-r)\psi~.
\label{4-9}
\een

For $ r < 1$ the condition \eqref{4-8} holds once \eqref{4-5} is obeyed. Therefore for $r < 1$ we always have, at late times 
\ben
(0 < r < 1,  \psi \gg t \gg 1) \qquad        \alpha_\pm \approx \pm \psi ~.
\label{4-20}
\een

When $1 < r < 2$ the condition \eqref{4-8} is not automatically obeyed once \eqref{4-5} is obeyed. We, therefore, need to consider the case $rt \gg (2-r)\psi$. It is then more useful to use a different power series expansion of the solution for $x^+$,
\ben
(x_-^{+})^{r-1} = 2 \cosh \psi ~e^{-rt} \left[1 + \sum_{n=1}^\infty \frac{c_n}{n!} (2 \cosh (\psi) )^{-n} \right]~,
\label{4-10}
\een
where the coefficients $c_n$ are given by
\begin{equation}
\begin{split}
c_n 
=&  (-1)^{n} \sum_{k=0}^{n} \binom{n}{k} \left( (2k-n) \tfrac{1}{r-1}\right)_{n-1} \left[ e^{-\frac{r}{r-1}t} (2\cosh(\psi))^{\frac{1}{r-1}} \right]^{2k-n} ~,
\end{split}
\label{4-10-2}
\end{equation}
according to the observations.
The $n \geq 1$ terms in the expansion can be ignored in the regime
\ben
\psi \gg t \gg \frac{2-r}{r} \psi~.
\label{4-11}
\een
In this case\footnote{Strictly speaking this requires $r$ to be close to 2. However, to get a qualitative insight, we will use this for the entire range.}
\ben
\log (x_-^{+})^{r-1} = -(r-1) \alpha_- = -rt + \log(2\cosh \psi)~.
\label{4-12}
\een
We therefore get
\begin{equation}
(1 < r < 2,  \psi \gg t \gg 1 )   \qquad \alpha_+ =  \psi , \quad \alpha_- = 
\begin{cases}
-\psi &  rt \ll (2-r) \psi \\
- \frac{\log (2 \cosh \psi) - rt }{r-1} & rt \gg (2-r) \psi \\
\end{cases}.
\label{w-soln}
\end{equation}
Finally consider $r > 2$. In this case the condition \eqref{4-8} is not possible, and \eqref{4-11} is always obeyed once \eqref{4-5} is obeyed. We therefore have
\begin{equation}
(r > 2, \psi \gg t \gg 1)  \qquad \alpha_+ =  \psi , \quad \alpha_- = - \frac{\log (2 \cosh \psi) - rt }{r-1}~.
\label{w-soln2}
\end{equation}

\section{Emergent space-time} \label{emergent-spt}

We can now use the results for $\alpha_\pm$ in the previous section to obtain the properties of the late time emergent space-time by constructing the Minkowsian coordinates defined in \eqref{2-14}. 

\subsection{Minkowskian coordinates near the turning point}

For the region near the turning point \eqref{smallxw2} leads to, at leading order, 
\begin{eqnarray}
q &=& \frac{1}{2} (\alpha_+ - \alpha_- ) = \left( \frac{ 2\delta x }{ (x_-^*)^{-1} + \frac{(r-1)(r-2)}{2} e^{rt} (x_-^*)^{r-1}  } \right)^{1/2}~, \nonumber \\
\tau &=& t + \log x_-^* \approx t + 
\begin{cases}
 \frac{\log (1-r) + rt}{2-r}~, & r<1 \\
0 ~, & r =1 \\
- \frac{\log (r-1) + rt}{r}~, & r>1 \\ 
\end{cases}=
\begin{cases}
 \frac{\log (1-r) + 2t}{2-r}~, & r<1 \\
t ~, & r =1 \\
- \frac{\log (r-1) }{r}~, & r>1 \\ 
\end{cases}~.
\label{5-5}
\end{eqnarray}
The expression for $\tau$ shows a clear difference between the cases $r \leq 1$ and $ r > 1$. In the former case, increasing $t$ leads to increasing $\tau$. However for $r > 1$ the Minkowskian time $\tau$ tends to a finite limit as $t \rightarrow \infty$.

\subsection{Minkowskian coordinates far from the turning point} \label{minkowskiancoord}

We now derive the expressions for the Minkowskian coordinates for the region far from the turning point.

\subsubsection{$ 0 < r < 1$}
These coordinates become, for $ 0 < r < 1$,
\ben
(0 < r < 1,  \psi \gg t \gg 1) \qquad   q=\psi~,\quad \tau = t~.
\label{5-1}
\een
The regime where this is valid is $\psi \gg t \gg 1$ which translates to $q \gg t \gg 1$.

\subsubsection{$1 < r < 2$} 
For $1 < r < 2$ and $\psi \gg t \gg 1$ \eqref{w-soln} leads to different expressions for $q, \tau$ in different regimes,
\ben
(1 < r < 2,  \psi \gg t \gg 1 ) \quad   q = 
\begin{cases}
\psi~, &  rt \ll (2-r) \psi \\
\frac{r}{2(r-1)} (\psi - t)~, & rt \gg (2-r) \psi \\
\end{cases}, \quad
\tau = \begin{cases}
t~, & rt \ll (2-r)\psi \\
\frac{(r-2)}{2(r-1)}(t - \psi)~, &  rt \gg (2-r) \psi
\end{cases}~.
\label{5-2}
\een
Re-expressing the regimes in terms of the corresponding $q$ we have
\ben
(1 < r < 2,  q , t \gg 1 ) \quad   q = 
\begin{cases}
\psi~, &  rt \ll (2-r) q \\
\frac{r}{2(r-1)} (\psi - t)~, & rt \gg (2-r) q\\
\end{cases}, \quad
\tau = \begin{cases}
t~, & rt \ll (2-r) q\\
\frac{(r-2)}{2(r-1)}(t - \psi), &  rt \gg (2-r) q
\end{cases}~.
\label{5-3}
\een

\subsubsection{$r > 2$}
Finally for $r > 2$ we have, for any $\psi \gg t \gg 1$ 
\begin{equation}
q = \frac{1}{2} \frac{r}{r-1} (\psi  -t )~,
\qquad
\tau  = \frac{1}{2} \frac{r- 2}{r-1} ( t - \psi )~.
\label{5-4}
\end{equation}
The regime is now characterized by $q , t \gg 1$.

\subsection{Global Properties of the emergent space-time}

The results above reveal a major difference between the cases $r < 1$ and $r > 1$. 

For $r < 1$, the global properties of the emergent space-time are similar to that obtained in the time independent case $r=0$. 

For $1 < r < 2$ the equation \eqref{5-5} shows that for small $q$ constant $t$ lines approach a fixed line $\tau = -\frac{1}{r}\log (r-1)$ as $t \rightarrow \infty$. For intermediate values of $q$ defined by the regime $t \gg \frac{2-r}{r} q \gg 1$ \eqref{5-3} shows that the $t \rightarrow \infty$ line approaches a space-like line
\ben
\tau = \frac{(2-r)}{r} q~.
\label{6-1}
\een
Finally for $\frac{2-r}{r} q   \gg  t \gg 1$ a constant large $t$ line approaches $\tau \rightarrow t$. These limiting lines lie {\em below} the line \eqref{6-1}.
Normally one would extend the space-time beyond this line. However the fundamental description of the theory is in terms of the matrix model, and the matrix model time $t$ is already infinite and the true time evolution of the system stops here. In other words, the emergent space-time perceived by excitations of the collective field theory has a space-like boundary defined by \eqref{6-1}.

For $ r > 2$ the behavior of constant $t$ lines for large $t$ is similar for small $q$, i.e. they approach a line $\tau = -\frac{1}{r}\log (r-1)$. For any $q , t \gg 1$ \eqref{5-4} shows that they asymptotically approach the space-like line
\ben
\tau = -\frac{(r-2)}{r} q~.
\label{6-2}
\een
Once again, this is a space-like boundary. 

\subsubsection{Exact numerical results}
These conclusions are corroborated by an exact numerical calculation of constant $t$ surfaces in the emergent $\tau - q$ space-time. We directly calculate the $\alpha_\pm$ by solving the first equation in \eqref{2-1} numerically and obtain the Minkowsian coordinates \eqref{2-14}. This is then used to plot constant $t$ lines for various values of $t$. {In addition, we accompany the $\tau - q$ plots for each regime of $r$ with their corresponding Penrose diagrams in which $\tau$ and  $q$ are mapped to Penrose diagram coordinates $X$ and $T$ through the relations
\ben
X = {1\over2}\big( \arctan(q-\tau) + \arctan(q+\tau) \big),\quad  T = {1\over2}\big( -\arctan(q-\tau) + \arctan(q+\tau) \big)
\een
where $\tau=0,~q=\pm\infty\implies T=0,~X=\pm{\pi\over2}$ corresponds to spatial infinity,  $\tau=\pm\infty,~q=0,\implies T=\pm{\pi\over2},~X=0$ corresponds to timelike infinity, and $\tau=\pm q,~ q= + \infty \left( \text{or} ~ - \infty \right), \implies T\pm X = {\pi\over2} \left( \text{or} ~ - {\pi\over2} \right)$ corresponds to future (or past) null or lightlike infinity}.

\begin{figure}[!h]
\centering
\includegraphics[width=4.0in]{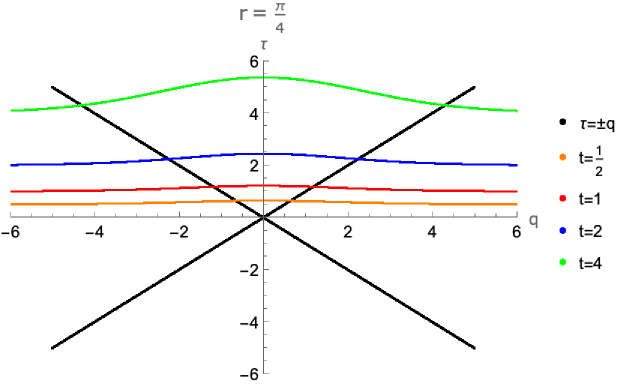}\\
\vspace{0.7cm}
\includegraphics[width=4.0in]{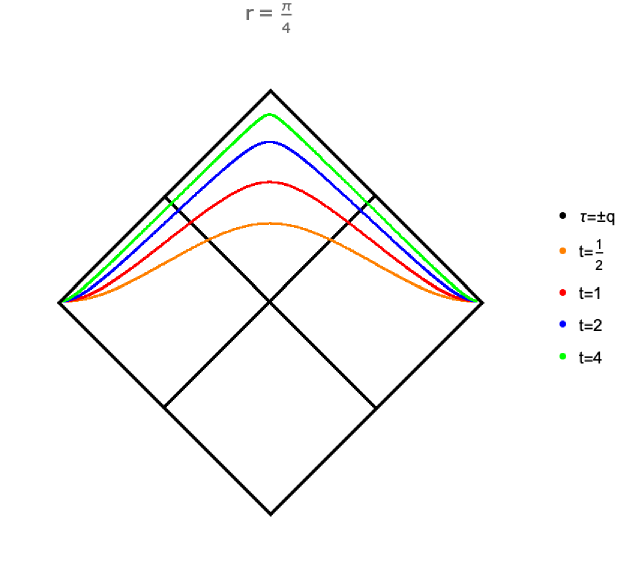}
\caption{Constant $t$ lines for $r = \frac{\pi}{4}$. Minkowski space (top) and Penrose diagram (bottom).}
\centering
\label{spacetimepi4}
\end{figure}

For $r < 1$ we indeed find that at late times constant $t$ lines approach constant $\tau$ lines and the emergent space-time is the full Minkowski space-time, consistent with \eqref{5-1}. Figure \ref{spacetimepi4} shows this for $r = \frac{\pi}{4}$.

\begin{figure}[!h]
\centering
\includegraphics[width=4.0in]{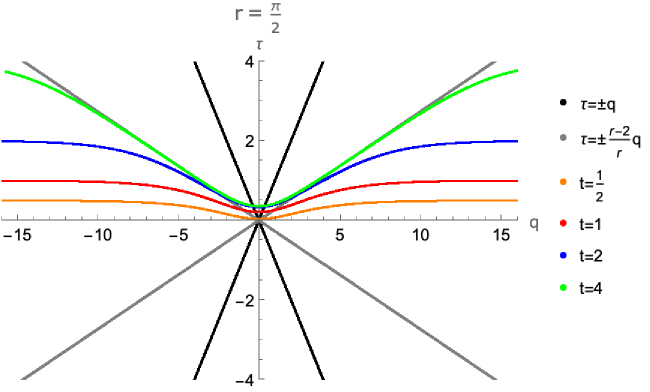}\\
\vspace{0.7cm}
\includegraphics[width=4.0in]{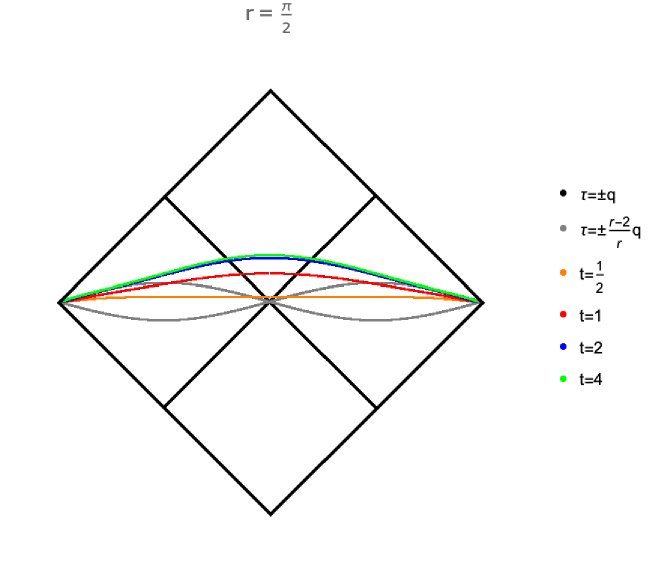}
\caption{Constant $t$ lines for $r = \frac{\pi}{2}$. Minkowski space (top) and Penrose diagram (bottom).}
\centering
\label{spacetimepi2}
\end{figure}

For $1 < r < 2$ the constant $t$ lines always lie below the space-like line $\tau = \frac{2-r}{r} q$. A constant $t$ line becomes tangential to this space-like boundary at intermediate values of $q$, deviating from it for large $q$. This is exactly what one expects from the discussion preceeding \eqref{6-1}. Note that the regime of $q$ over which the line \eqref{6-1} agrees with a constant $t$ line moves to larger and larger $q$ as $t$ increases. However a constant $t$ line is always below \eqref{6-1}. The latter therefore is a space-like boundary. Figure \ref{spacetimepi2} shows a typical result for $r = \pi/2$.

\begin{figure}[!h]
\centering
\includegraphics[width=4.0in]{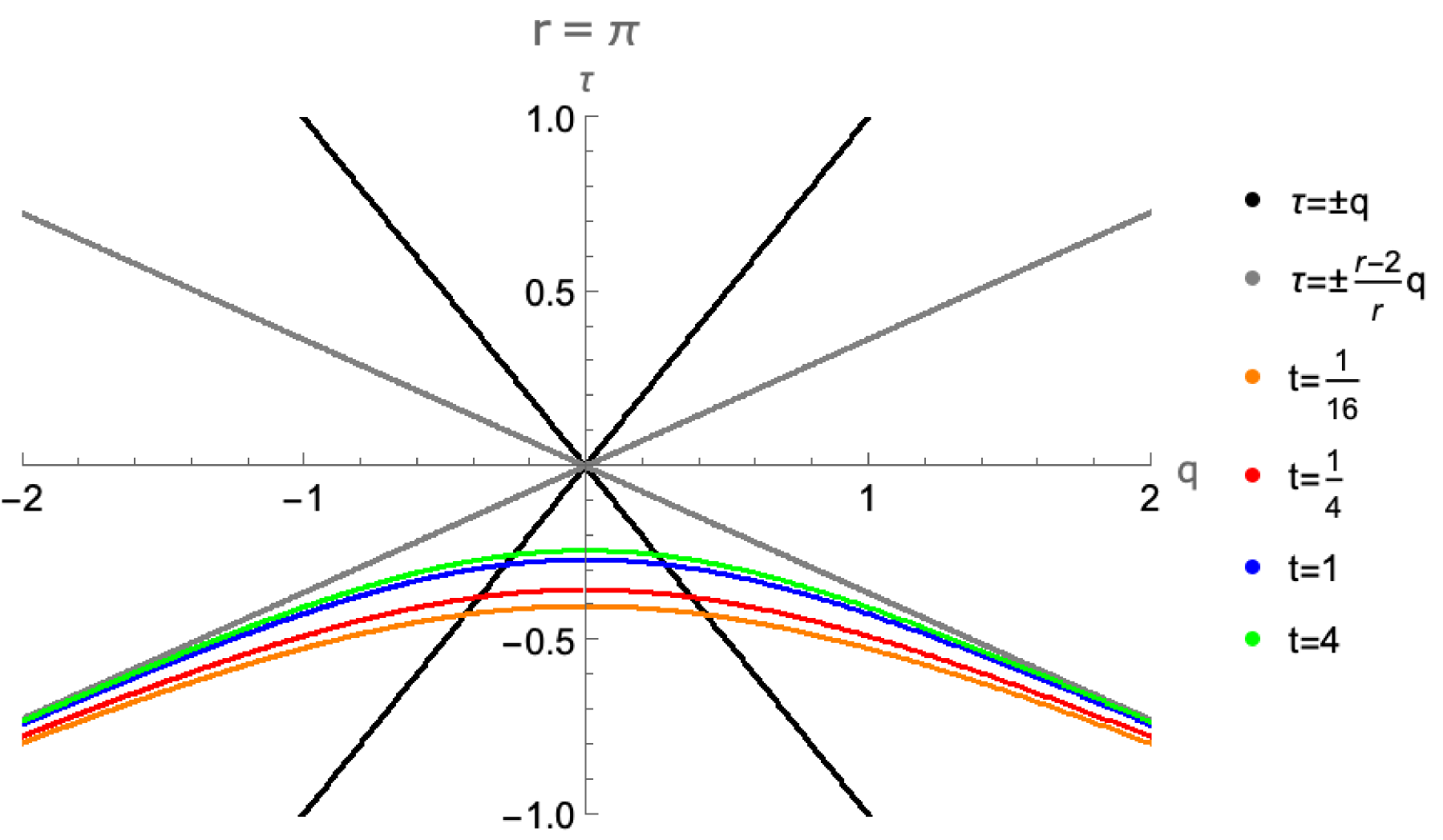}\\
\vspace{0.7cm}
\includegraphics[width=4.0in]{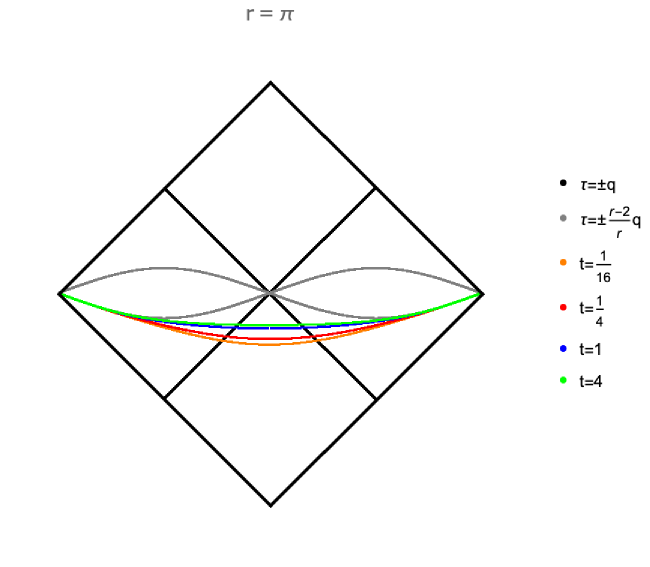}
\caption{Constant $t$ lines for $r = \pi$. Minkowski space (top) and Penrose diagram (bottom).}
\centering
\label{spacetimepi}
\end{figure}

When $r > 2$ constant $t$ lines are again always below the spacelike line
(\ref{6-2}), approaching it for large $q$, in agreement with our analysis above.
Figure \ref{spacetimepi} shows the result for $r = \pi$.

\section{Divergent Couplings at the boundary} \label{divcp}

We have seen that for $r > 1$ there is a space-like boundary for the emergent space-time perceived by excitations of the collective field. In this section we will compute the cubic couplings which appear in the collective field action \eqref{2-12} and show that at least one of the couplings diverges along this space-like boundary. 

The cubic couplings are obtained by first evaluating $P_\pm (x,t) = p(\alpha_\pm, t)$ where $p(\alpha_\pm,t)$ is obtained by inserting $\alpha_\pm$ in \eqref{2-1} 
These can then be used to calculate the quantities 
$(\partial_x \varphi_0)$ and $(\partial_t \varphi_0)/(\partial_x \varphi_0)$ using \eqref{1-8a}. The latter directly provide the cubic couplings in \eqref{2-12}.

The couplings are of course expected to be large near the turning point of the Fermi surface at all times, as for the static solution: roughly speaking this is the Liouville wall. We therefore estimate these couplings in the asymptotic regions far from the turning point. 

\begin{itemize}

\item{} For $0 < r < 1$ and in the region $ q \gg t \gg 1$ we get, using \eqref{4-20} and \eqref{5-1}
\begin{equation}
\frac{1}{(\partial_x \varphi_0)^2} \sim e^{-2q},\qquad 
\frac{\partial_t \varphi_0}{(\partial_x \varphi_0)^3} 
\sim e^{r(t -q) -q}, \qquad 
 \frac{(\partial_t \varphi_0)^2}{(\partial_x \varphi_0)^4} \sim  e^{2r(t-q)} ~.
 \label{7-1}
\end{equation}
In the region of space-time of interest all of these couplings become small.

\item{} For $1 < r < 2$ we have two cases in the region $ q, t \gg 1$

\subitem{(i)} When $rt \ll (2-r)q$ we need to use the first line of \eqref{w-soln} and the first line of \eqref{5-2}. The couplings evaluate to 
\begin{equation}
\frac{1}{(\partial_x \varphi_0)^2} \sim e^{-2q},\qquad 
\frac{\partial_t \varphi_0}{(\partial_x \varphi_0)^3} 
 \sim e^{rt - (2-r)q -q}, \qquad 
 \frac{(\partial_t \varphi_0)^2}{(\partial_x \varphi_0)^4} \sim  e^{2(rt - (2-r)q )} ~.
 \label{7-2}
\end{equation}
None of these couplings diverge.

\subitem{(ii)} When $rt \gg (2-r)q$ we need to use the second lines of \eqref{w-soln} and \eqref{5-2}.
In this case the first two of the three couplings are
\begin{equation}
\frac{1}{(\partial_x \varphi_0)^2} \sim 
 \left[  \frac{1}{2} e^{ t  - \frac{2(r-1)^2}{r} q } + \frac{1}{2} e^{ -t + \frac{2}{r}q }  \right]^{-2} ~, \qquad
\frac{\partial_t \varphi_0}{(\partial_x \varphi_0)^3} \sim \frac{ e^{ t  + \frac{2(r-1)}{r} q } }{ \left[ \frac{1}{2} e^{ t  - \frac{2(r-1)^2}{r} q } + \frac{1}{2} e^{ -t + \frac{2}{r}q }\right]^{2} }~.
\label{7-3}
\end{equation}
These two couplings may or may not be divergent depending on $(r,q)$. The third cubic coupling becomes
\begin{equation}
 \frac{(\partial_t \varphi_0)^2}{(\partial_x \varphi_0)^4}  \sim  \frac{ e^{ 2( t  + \frac{2(r-1)}{r} q ) } }{ \left[ \frac{1}{2} e^{ t  - \frac{2(r-1)^2}{r} q } + \frac{1}{2} e^{ -t + \frac{2}{r}q } \right]^{2} }~.
 \label{7-4}
\end{equation}
This coupling always diverge at late times. Note that this is the regime of $q$ where a constant-$t$ line at large $t$ agrees with the space-like line given by \eqref{6-1}. Therefore one of the couplings always diverges along the entire space-like line \eqref{6-1}, which is the boundary of the emergent space-time.

\item{} When $ r > 2$ we need to use \eqref{w-soln2} and \eqref{5-4}. The couplings now become
\begin{equation}
\frac{1}{(\partial_x \varphi_0)^2} \sim 
 \left[ \sinh \left( t - \frac{2}{r}q \right) \right]^{-2} ~, \qquad
\frac{\partial_t \varphi_0}{(\partial_x \varphi_0)^3}  \sim \frac{ e^{ t  + \frac{2(r-1)}{r} q } }{ \left[ \sinh \left( t - \frac{2}{r}q \right)\right]^{2} }
\label{7-5}
\end{equation}
and 
\begin{equation}
 \frac{(\partial_t \varphi_0)^2}{(\partial_x \varphi_0)^4}  \sim  \frac{ e^{ 2( t  + \frac{2(r-1)}{r} q ) } }{ \left[ \sinh \left( t - \frac{2}{r}q \right) \right]^{2} }~.
\label{7-6}
\end{equation}
In this case a constant $t$ line for $q, t \gg 1$ is described by the space-like boundary \eqref{6-2}. It may be checked that the coupling \eqref{7-6} is always divergent all along this line.

\end{itemize}

We therefore conclude that whenever the emergent space-time has a space-like boundary a cubic coupling is divergent on the boundary.


\section{Discussion} \label{discuss}

In the examples discussed above, perturbative collective field theory breaks down at late times -- this appears along a space-like line in the emergent space-time. This is akin to a space-like singularity in the sense of singularities of the low energy description. The fermion time evolution remains well defined and should be calculable along the lines of \cite{ds}. What this really means that the space-time interpretation of the model, in which the collective excitations become fields in a $1+1$ dimensional space-time requires modification in this region.

The time dependent Fermi surfaces we considered, which correspond to vertex operator deformations on the worldsheet, are, however, somewhat pathological. The filled regions of the fermion phase space gets pushed into the asymptotic region at late times. It appears that the fermions are leaking out to infinity. For $r > 1$ we found that at the same time the collective field couplings become large. This, however, cannot happen if the number of fermions, $N$, is finite, however large. 

When $N$ is finite, the potential cannot be an inverted oscillator potential which appears in the double scaled limit where $N \rightarrow \infty$ keeping $g_s$ fixed. Rather the modifications of the potential in the regions far from the maximum now become important. While the precise nature of this modification is not universal, we may try to understand what happens to the above picture when we put in a hard wall at a location $O(\sqrt{N})$. Now these fermions will get reflected from the wall and will modify the profiles of the Fermi surfaces substantially. These modifications can be understood in a controlled fashion where time dependences of the type we consider in this paper arise from a quantum quench. In this latter case, disconnected pieces of the Fermi sea will appear from the wall, leading to non-quadratic profiles. (See Figure 12 and Appendix D of \cite{dhl}). One also needs to address the consistency issues raised in \cite{consistency}. 

It is significant that to understand the late time physics we not only need finite string coupling effects, but finite $N$ effects as well. Note that these are related, but distinct effects, since the double scaled coupling is finite even at $N=\infty$. There are several phenomena in Matrix models and String Theory where finite $N$ effects play key roles. For example, in the free single matrix problem, collective variables incorporating the trace relations are necessary to reproduce the exact fermion dispersion relation \cite{jevicki-exact,jevicki-exact2}. In $AdS/CFT$ these collective variables account for the physics of giant gravitons \cite{corley} and the Stringy Exclusion Principle \cite{exclusion}. The finiteness of the entanglement entropy of a region of the eigenvalue space in the single matrix problem obtained in the fermionic formulation \cite{das,hartnoll} finds an explanation in a bosonic theory of these finite $N$ collective variables \cite{djz}.  Alternatively, a bosonization procedure for a {\em finite} number of fermions \cite{gautam} makes this finiteness manifest. More recently, it has been found that finite $N$ trace relations in multi-matrix models lead to rather drastic consequences about the nature of the Hilbert space \cite{jr1} \footnote{There are similar issues in vector model holography. See \cite{rdmk} for a discussion.}.

There are several possible ways to proceed further. One possibility is to develop the formulation of \cite{jevicki-exact, jevicki-exact2}, taking into account the IR wall. Perhaps the collective theory of $q$-oscillators \cite{qoscillator} as a way to incorporate finite $N$ effects will be useful in this. 
A second possibility is to use the full {\em quantum} theory of phase space density \cite{dmw,son}. We plan to explore these approaches in the near future.

Finally we would like to explore other types of time dependent solutions in this model. In BFSS of IKKT matrix models \cite{bran,bran2} find solutions which correspond to an expanding universe starting from a non-geometric phase. These would be somewhat similar to the time reversed version of our backgrounds (i.e. corresponding to $\lambda_+= 0, \lambda_- \neq 0$. It would be interesting to explore such backgrounds in  the $c=1$ model.

\section*{Acknowledgements} S.R.D would like to thank A. Jevicki, G. Mandal and T. Takayanagi for discussions at various stages of this work. An early version of this work was presented at the workshop ``Quantum Gravity and Information in an Expanding Universe'' at Yukawa Institute for Theoretical Physics. The work of S.R.D. is partially supported by National Science Foundation grants NSF-PHY/211673 and NSFPHY/2410647, and a Jack and Linda Gill Chair Professorship. The work of S.D.H is supported by KIAS Grant PG096301.

\appendix

\section{The two intersections of Fermi surface and constant-$x$ line when $\lambda_+ <0,~\lambda_-=0$} \label{app-a}

In this appendix we calculate the two solutions of \eqref{4-1} (or equivalently \eqref{1-3}), $x_-^{\pm}$. The solutions are valid for all positive $r$. Our strategy is to assume that the $r$-th order term in $x_-$ in \eqref{4-1} is a correction to the quadratic equation of $x_-$, and therefore the two exact solutions are power series' in $e^{rt}$. Obviously, when $e^{rt}$ becomes large there will be some problems with the convergence of the two series'. We will discuss the convergence of $x_-^{\pm}$ in appendix \ref{app-a-1} and introduce better power series forms to express $x_-^+$ in appendix \ref{app-a-2}.

According to our assumption, the exact solution, $x_-^{\pm}$, can be expressed by the Taylor series
\begin{equation}
x_-^{\pm} = e^{\pm \psi} \left( 1+ \sum_{n=1}^{\infty} \frac{a_n^{\pm}}{n!} e^{n\cdot rt}  \right)~, 
\label{bellsoln1}
\end{equation}
where the $a_n^{\pm}$'s are a series of functions of $\psi$ and $t$ that need to be determined. Utilising Fa\`a di Bruno's formula, the $r$-order term in $x_-$ is therefore
\begin{equation}
\left( x_-^{\pm} \right)^r = e^{\pm r \psi} \left[ 1 + \sum _{n=1}^{\infty }{\frac {\sum _{k=1}^{n} (r)_k B_{n,k}(a_{1}^{\pm},\dots ,a_{n-k+1}^{\pm})}{n!}} e^{n\cdot rt} \right]~,
\label{a-2}
\end{equation}
where $(r)_n$ is the falling factorial
\begin{equation}
\nonumber
(r)_{n}=\underbrace {r(r-1)(r-2)\cdots (r-n+1)}_{n{\text{ factors}}}~,
\end{equation}
and $B_{n,k}(a_{1}^{\pm},\dots ,a_{n-k+1}^{\pm})$ represents the partial exponential Bell polynomial
\begin{equation}
B_{n,k}(x_{1},x_{2},\dots ,x_{n-k+1})=n!\sum_{j_{1}+j_{2}+\cdots +j_{n-k+1}=k, \atop  j_{1}+2j_{2}+3j_{3}+\cdots +(n-k+1)j_{n-k+1}=n} \prod _{i=1}^{n-k+1}{\frac {x_{i}^{j_{i}}}{(i!)^{j_{i}}j_{i}!}} 
\end{equation}
with conventions $B_{0,0}=1$, $B_{n,0}=0$ (for $n\geq 1$) and $B_{0,k}=0$ (for $k\geq 1$).
In particular, when $r=2$, \eqref{a-2} can be further simplified into 
\begin{equation}
\left( x_-^{\pm} \right)^2 = e^{\pm 2 \psi} \left[ 1 + 2 \sum _{n=1}^{\infty }{\frac {a_{n}^{\pm}}{n!}} e^{n\cdot rt} + \sum _{n=1}^{\infty }{\frac {  \sum_{k=1}^{n-1} \binom{n}{k} a_k^{\pm} a_{n-k}^{\pm}}{n!}} e^{n\cdot rt} \right]~,
\label{a-3}
\end{equation}
Inserting \eqref{a-2} and \eqref{a-3} into \eqref{4-1}, we obtain
\begin{equation}
\begin{split}
  e^{\pm 2 \psi}   \sum _{n=1}^{\infty }{\frac {  \sum_{k=1}^{n-1} \binom{n}{k} a_k^{\pm} a_{n-k}^{\pm}}{n!}} e^{n\cdot rt}  & + \left( e^{\pm 2 \psi}  - 1  \right)   \sum_{n=1}^{\infty} \frac{a_n^{\pm}}{n!} e^{n\cdot rt} \\
  + e^{rt}e^{\pm r \psi} &  \sum _{n=0}^{\infty }{\frac {\sum _{k=0}^{n} (r)_k B_{n,k}(a_{1}^{\pm},\dots ,a_{n-k+1}^{\pm})}{n!}} e^{n\cdot rt} = 0 ~,\\
\end{split}
\end{equation}
which leads to the recurrence relation between the $a_n^{\pm}$'s
\begin{equation}
a_n^{\pm} =\frac{ e^{\pm 2 \psi} }{1 -  e^{\pm 2 \psi} }\cdot  \sum_{k=1}^{n-1} \binom{n}{k} a_k^{\pm} a_{n-k}^{\pm} + \frac{ e^{\pm r \psi} }{1 -  e^{\pm 2 \psi} } \cdot n \sum _{k=0}^{n-1} (r)_k B_{n-1,k}(a_{1}^{\pm},\dots ,a_{n-k}^{\pm})~, \qquad n \ge 1.
\label{a-5}
\end{equation}

It will be more convenient to introduce another series of functions of $\psi$ and $t$, $b_n^{\pm}$, such that
\begin{equation}
a_n^{\pm} = \left[ \frac{ e^{\pm r \psi} }{1 -  e^{\pm 2 \psi} } \right]^n b_n^{\pm}~.
\end{equation}
The advantage of the $b_n^{\pm}$s are that the recurrence relation \eqref{a-5} now becomes
\begin{equation}
b_n^{\pm} =\frac{ e^{\pm 2 \psi} }{1 -  e^{\pm 2 \psi} }\cdot  \sum_{k=1}^{n-1} \binom{n}{k} b_k^{\pm} b_{n-k}^{\pm} +  n \sum _{k=0}^{n-1} (r)_k B_{n-1,k}(b_{1}^{\pm},\dots ,b_{n-k}^{\pm})~, \qquad n \ge 1,
\label{a-7}
\end{equation}
where we have utilised the property of Bell polynomials
\begin{equation}
 B_{n,k}(\alpha \beta x_{1},\alpha \beta ^{2}x_{2},\ldots ,\alpha \beta ^{n-k+1}x_{n-k+1})=\alpha ^{k}\beta ^{n}B_{n,k}(x_{1},x_{2},\ldots ,x_{n-k+1})~.
\end{equation}
\eqref{a-7} implies that $b_n$s are polynomials of $\frac{ e^{\pm 2 \psi} }{1 -  e^{\pm 2 \psi} } $.
Let $y_{\pm} = \frac{ e^{\pm 2 \psi} }{1 -  e^{\pm 2 \psi} } $, the first five $b_n^{\pm}$s are
\begin{eqnarray}
b_1^{\pm} &=& 1 ~,\nonumber \\
b_2^{\pm} &=& 2 y_{\pm} + 2r ~,\nonumber  \\
b_3^{\pm} &=&  12 y_{\pm}^2 + 6 \cdot 3 r  y_{\pm} + 3r (3r-1) ~,\nonumber  \\
b_4^{\pm} 
&=&120 y_{\pm}^3+60 \cdot 4r y_{\pm}^2 + 12 \cdot 4 r(4r -1) y_{\pm} + 4r (4r-1)(4r-2)  ~,\nonumber \\
b_5^{\pm} 
&=& 1680 y_{\pm}^4 +840 \cdot 5 r y_{\pm}^3+ 180 \cdot 5r (5r-1) y_{\pm}^2+ 20 \cdot 5r (5r-1)(5r-2) y_{\pm} \nonumber \\
&& +5r(5r-1)(5r-2)(5r-3) ~.
\end{eqnarray}
An observation is therefore
\begin{equation}
b_n^{\pm} =  \sum_{k=1}^{n} \frac{ (n +k -2 )! }{(k-1)! (n-k)!} (nr)_{n-k} \left[  \frac{ e^{\pm 2 \psi} }{1 -  e^{\pm 2 \psi} } \right]^{k-1} ~.
\tag{\ref{4-3}}
\end{equation}

In summary, the two exact solutions of \eqref{4-1} are
\begin{equation}
x_-^{\pm} = e^{\pm \psi} \left[ 1+ \sum_{n=1}^{\infty} \frac{1}{n!} \left( \frac{ e^{ r ( t \pm  \psi )} }{1 -  e^{\pm 2 \psi} } \right)^n  \sum_{k=1}^{n} \frac{ (n +k -2 )! }{(k-1)! (n-k)!} (nr)_{n-k} \left(  \frac{ e^{\pm 2 \psi} }{1 -  e^{\pm 2 \psi} } \right)^{k-1} \right]~. 
\label{a-9}
\end{equation}

\subsection{Convergence of the series} \label{app-a-1}

In our main text, we are more interested in late time behaviour, which brings us to the convergence problems mentioned at the very beginning of this section. Below we check the convergence of the series expression of $x_-^{\pm}$ in the regime
\begin{equation}
\psi \gg t \gg 1~.
\tag{\ref{4-5}}
\end{equation}

The convergence of the series $x_-^-$ in \eqref{a-9} is relatively easy to check. Notice that since
\begin{equation}
 \frac{ e^{- d \psi} }{1 -  e^{- 2 \psi} } \approx e^{- d\psi} \ll 1~,\quad d = 2, r~,
\end{equation}
we only have to keep the $k=1$ term in $b_n^{-}$ obtaining
\begin{equation}
b_n^{-} \approx (nr)_{n-1} ={\frac {\Gamma ( rn+1)}{\Gamma ( rn -n+2)}}~.
\end{equation}
where $\Gamma (x)$ is the Gamma function. 
This simplifies \eqref{a-9} into
\begin{equation}
x_-^{-} = e^{ - \psi} \left[ 1+ \sum_{n=1}^{\infty} \frac{1}{n!} {\frac {\Gamma ( rn +1)}{\Gamma ( rn -n+2)}} e^{ n \cdot r ( t -  \psi ) } \right]~.
\end{equation}
The series (absolutely) converges if
\begin{equation}
e^{   r ( t -  \psi ) } < 
\lim_{n \to \infty}  n \left|   {\frac {\Gamma ( r(n-1) +1)}{\Gamma ( (r-1)(n-1)+2)}}  {\frac{\Gamma ( (r-1)n+2)} {\Gamma ( r n+1)}}   \right|~,
\label{a-15}
\end{equation}
according to the ratio test of convergence. The R.H.S. of \eqref{a-15} can be evaluated by Gauss’s multiplication formula
\begin{equation}
\Gamma\left(nz\right)=(2\pi)^{(1-n)/2}n^{nz-(1/2)}\prod_{k=0}^{n-1}\Gamma\left(z+\frac{k}{n}\right), \quad nz \ne 0, -1, -2, -3, \cdots,
\label{a-16}
\end{equation}
from which we learn that\footnote{Notice that here we consider everything in the limit $n \to \infty$, where $rn$ is positive while $(r-1)n$ can be non-positive when $r \in (0,1]$. But in general $(r-1)n$ is not a non-positive integer when $n \to \infty$, unless $r=1$. }
\begin{equation}
\begin{split}
{\frac {\Gamma ( rn +1)}{\Gamma ( rn -n+2)}} 
=& \frac{r}{(r-1) [ (r-1)n +1 ]} n^n \prod_{k=0}^{n-1} \left( r -1 +\frac{k}{n} \right) ~, \quad r \ne 1
\end{split}
\label{a-17}
\end{equation}
and therefore
\begin{equation}
\begin{split}
e^{   r ( t -  \psi ) } < & \lim_{n \to \infty}  n \left|   {\frac {\Gamma ( r(n-1) +1)}{\Gamma ( (r-1)(n-1)+2)}}  {\frac{\Gamma ( (r-1)n+2)} {\Gamma ( r n+1)}}   \right| \\
=&  \lim_{n \to \infty} \left| \frac{ (r-1)n +1}{ (r-1)(n-1) +1 } \left( \frac{n-1}{n} \right)^{n-1} \prod_{k=0}^{n-2} \frac{ r -1 +\frac{k}{n-1}}{r -1 +\frac{k}{n}} \cdot \left( r - 1 +\frac{n-1}{n} \right)^{-1} \right| \\
=& \frac{1}{r}~.\\
\end{split}
\label{a-18}
\end{equation}
The inequality is valid as well when $r=1$, in which case ${\frac {\Gamma ( rn +1)}{\Gamma ( rn -n+2)}} = n!$ and the ratio test of convergence shows $e^{t-\psi} < 1$.
Equivalently, the convergence of series $x_-^-$ requires
\begin{equation}
\psi - t > \frac{1}{r} \log (r) 
\label{a-19}
\end{equation}
for arbitrary positive $r$.
Noting that $\frac{1}{r} \log (r)$ has a maximum $e^{-1}$ at $r = e$, \eqref{a-19} is satisfied in the regime of interest \eqref{4-5} and therefore the series $x_-^-$ converges. 

Now consider the convergence of the series $x_-^+$ in \eqref{a-9}. Again we first simplify the series in the regime \eqref{4-5}. Since
\begin{equation}
 \frac{ e^{ d \psi} }{1 -  e^{2 \psi} }  \approx - e^{(d-2)\psi}  ~,\quad d = 2, r~,
\end{equation}
$b_n^{+}$ can be expressed by Gamma functions
\begin{equation}
b_n^{+} \approx  \sum_{k=1}^{n} \frac{ (n +k -2 )! }{(k-1)! (n-k)!} (nr)_{n-k} (-1)^{k-1} = \frac{\Gamma (r n-n+1)}{\Gamma (r n-2 n+2)}
 ~.
\end{equation}
The series $x_-^+$ now becomes
\begin{equation}
x_-^{+} = e^{ \psi} \left[ 1+ \sum_{n=1}^{\infty} \frac{1}{n!}  \frac{\Gamma (r n-n+1)}{\Gamma (r n-2 n+2)}  \left( -  e^{ r t + (r-2) \psi }  \right)^n \right]~. 
\label{a-22}
\end{equation}
It is absolutely convergent if
\begin{equation}
\begin{split}
 e^{ r t + (r-2) \psi }  < & \lim_{n \to \infty} n \left| \frac{\Gamma ( (r-1) (n-1)+1)}{\Gamma ( (r-2) (n-1) +2)} \frac{\Gamma ( (r-2) n +2)}{\Gamma ( (r-1) n+1)}  \right|  \\
 =& \lim_{n \to \infty} \left| \frac{ (r-2)n +1}{ (r-2)(n-1) +1 } \left( \frac{n-1}{n} \right)^{n-1} \prod_{k=0}^{n-2} \frac{ r -2 +\frac{k}{n-1}}{r -2 +\frac{k}{n}} \cdot \left( r - 2 +\frac{n-1}{n} \right)^{-1} \right| \\
=& \frac{1}{|r-1|}~, \quad \text{for} ~ r \ne 1, 2,
\end{split}
\label{a-23}
\end{equation}
based on a similar evaluation to $x_-^-$ in \eqref{a-17} and \eqref{a-18}. The inequality is still valid when $r=2$, in which case $\frac{\Gamma (r n-n+1)}{\Gamma (r n-2 n+2)} =n!$ and the ratio test shows $e^{  2t  }  < 1$. \eqref{a-23} is equivalent to the condition
\begin{equation}
 \frac{2-r}{r}\psi  -  t > \frac{1}{r} \log | r -1 |~
\label{a-24}
\end{equation}
when $r >0$ and $r \ne 1$. The R.H.S. of \eqref{a-24}, $\frac{1}{r} \log | r -1 |$, has a maximum value of approximately $0.278465$, which implies that the series $x_-^+$ always absolutely converges when
\begin{equation}
 \frac{2-r}{r}\psi  \gg  t ~.
\tag{\ref{4-8}'}
\end{equation}
On the other hand, for $r=1$, from $\frac{\Gamma (r n-n+1)}{\Gamma (r n-2 n+2)} = \Gamma(2-n)^{-1}$ we obtain the radius of convergence
\begin{equation}
e^{t-\psi} < \lim_{n \to \infty} n \left| \frac{\Gamma ( 2-n )}{\Gamma (2-(n-1))}  \right| =  \lim_{n \to \infty} \left| \frac{n}{2-n} \right|=1~,
\end{equation}
or equivalently $\psi > t$. Hence \eqref{4-8} is also the sufficient condition of the absolute convergence of $x_-^+$ at $r=1$.  

It should be recognised that the regime given in \eqref{4-8} is not equivalent to the regime of interest \eqref{4-5}: when $r \le 1$, the satisfaction of \eqref{4-5} ensures that of \eqref{4-8} while when $1<r <2 $, the regime in \eqref{4-5} can be further divided into two regimes, one described by \eqref{4-8} and the other by \eqref{4-11}. As for $r \ge 2$, \eqref{4-8} can never be satisfied within the regime \eqref{4-5}. This is how we classify the solutions in section \ref{solnfar}.

As it is the end of this discussion, we make one more comment on the convergence of series $x_-^+$. So far we have concentrated on the absolute convergence of the series expansion of $x_-^{\pm}$ in \eqref{a-9}, but this may not reflect the exact convergence of the series $x_-^+$, since $x_-^+$ can conditionally converge due to the extra negative sign in the even terms, rotating the variable of the power series from $e^{rt + (r-2) \psi}$ to $-e^{rt + (r-2) \psi}$. However, we will not continue checking the (conditional) convergence of $x_-^+$ -- because when \eqref{4-8} is far from being satisfied, it is difficult to see straightforwardly the asymptotic behaviour of $x_-^+$ that we are interested in. Instead, in this case we will look for a better series form for $x_-^+$ in the next section.

\subsection{Another series expansion of $x_-^+$} \label{app-a-2}

Below we look for a better power series expression of the solution to \eqref{1-3} (or equivalently \eqref{4-1}), $x_-^+$, when $\psi$ (or $x$) and $t$ satisfy 
\begin{equation}
\psi \gg t \gg \frac{2-r}{r} \psi
\tag{\ref{4-11}}
\end{equation}
rather than \eqref{4-8}. Note that such a regime is only meaningful when $r>1$: when $r>2$, \eqref{4-11} is satisfied as long as we consider the regime \eqref{4-5}; when $1<r<2$, \eqref{4-11} and \eqref{4-8} are the two typical regimes within the region of interest \eqref{4-5}.

We start from \eqref{1-3}. Since $r>1$, we can introduce a variable $y$ such that
\begin{equation}
(x_-^+)^{r-1} = e^{-rt} \cdot (2x) \cdot y ~.
\end{equation}
In terms of $y$, \eqref{1-3} becomes
\begin{equation}
y + e^{-\frac{r}{r-1}t} (2\cosh (\psi) )^{\frac{2-r}{r-1}} y^{\frac{1}{r-1}} + e^{\frac{r}{r-1}t} (2 \cosh (\psi) )^{-\frac{r}{r-1}} y^{-\frac{1}{r-1}} -1 =0~.
\label{a-27}
\end{equation}
In the regime \eqref{4-11}, the factors of the second and the third terms of \eqref{a-27}, i.e. $e^{-\frac{r}{r-1}t} (2\cosh (\psi) )^{\frac{2-r}{r-1}}$ and $ e^{\frac{r}{r-1}t} (2 \cosh (\psi) )^{-\frac{r}{r-1}}$, are much smaller than 1, which implies that the second and the third terms of \eqref{a-27} can be regarded as corrections to the linear equation of $y$. 

In this spirit, we assume
\begin{equation}
y = 1 + \sum_{n=1}^{\infty} \frac{c_n}{n!} \left( 2 \cosh (\psi) \right)^{-n}~,
\end{equation}
and therefore
\begin{equation}
(x_-^{+})^{r-1}  = 2 \cosh (\psi) e^{-rt} \left[ 1 + \sum_{n=1}^{\infty} \frac{c_n}{n!} \left( 2 \cosh (\psi) \right)^{-n} \right]~,
\tag{\ref{4-10}}
\end{equation}
where $c_n$s are a series of functions of $\psi$ and $t$ that need to be determined. 
As done at the beginning of appendix \ref{app-a}, by utilising Fa\`a di Bruno's formula, we can express the $\pm \frac{1}{r-1}$ terms in $y$ in \eqref{a-27} as the following series
\begin{equation}
\begin{split}
y^{\pm \frac{1}{r-1}} 
=& 1 + \sum_{n=1}^{\infty} \frac{\sum_{k=1}^n \left(\pm \frac{1}{r-1}\right)_k B_{n,k}(c_1, \cdots, c_{n-k+1})}{n!} \left( 2 \cosh (\psi) \right)^{-n}
\end{split}
\label{a-29}
\end{equation}
Inserting \eqref{a-29} into \eqref{a-27} we obtain
\begin{equation}
\begin{split}
  \sum_{n=1}^{\infty} \frac{c_{n}}{n !} \left( 2 \cosh (\psi) \right)^{-n}  \qquad \qquad & \\
  + \sum_{n=0}^{\infty} \frac{\left( 2 \cosh (\psi) \right)^{-n-1}}{n!}   \times   \sum_{k=0}^{n} & \left[\left( \tfrac{1}{r-1}\right)_k e^{-\frac{r}{r-1}t} (2\cosh(\psi))^{\frac{1}{r-1}} \right. \\
 & \left. +\left( - \tfrac{1}{r-1}\right)_k e^{\frac{r}{r-1}t} (2\cosh(\psi))^{-\frac{1}{r-1}}  \right] B_{n,k}(c_1, \cdots, c_{n-k+1}) =0~.
\end{split}
\end{equation}
The resulting recurrence relation is 
\begin{eqnarray}
c_1 &=& - e^{-\frac{r}{r-1}t} (2\cosh(\psi))^{\frac{1}{r-1}} - e^{\frac{r}{r-1}t} (2\cosh(\psi))^{-\frac{1}{r-1}}  \nonumber \\
c_{n} &=& - n \sum_{k=0}^{n-1} \left[\left( \tfrac{1}{r-1}\right)_k e^{-\frac{r}{r-1}t} (2\cosh(\psi))^{\frac{1}{r-1}} +\left( - \tfrac{1}{r-1}\right)_k e^{\frac{r}{r-1}t} (2\cosh(\psi))^{-\frac{1}{r-1}}  \right] \nonumber \\
&& \qquad \qquad \times B_{n-1,k}(c_1, \cdots, c_{n-k}), 
\qquad  \qquad \qquad \qquad \qquad  \qquad \qquad  n \ge 2~. 
\end{eqnarray}
We calculate the first five $c_n$s (let $\beta = e^{-\frac{r}{r-1}t} (2\cosh(\psi))^{\frac{1}{r-1}}$ ), which are
\begin{eqnarray}
c_1 &=& - \left( - \tfrac{1}{r-1}\right)_0 \beta^{-1} - \left(  \tfrac{1}{r-1}\right)_1\beta ~, \nonumber\\
c_2  &=&  \left( -2 \tfrac{1}{r-1}\right)_1\beta^{-2}+ \left( 2 \tfrac{1}{r-1}\right)_1\beta^2~, \nonumber\\
c_3  & =&  -\left( -3\tfrac{1}{r-1}\right)_2 \beta^{-3} - 3 \left( -\tfrac{1}{r-1}\right)_2 \beta^{-1} - 3 \left( \tfrac{1}{r-1}\right)_2 \beta -  \left( 3\tfrac{1}{r-1}\right)_2 \beta^{3}~,   \nonumber \\
c_4 &=&  \left( -4\tfrac{1}{r-1}\right)_3 \beta^{-4} + 4 \left( -2\tfrac{1}{r-1}\right)_3 \beta^{-2} + 4 \left( 2\tfrac{1}{r-1}\right)_3 \beta^2 +  \left( 4\tfrac{1}{r-1}\right)_3 \beta^{4}~, \nonumber \\
c_5
&=& - \left( -5\tfrac{1}{r-1}\right)_4 \beta^{-5} - 5 \left( -3\tfrac{1}{r-1}\right)_4 \beta^{-3} - 10 \left( -\tfrac{1}{r-1}\right)_4 \beta^{-1} \nonumber \\
&& -  10 \left( \tfrac{1}{r-1}\right)_4 \beta -5 \left( 3\tfrac{1}{r-1}\right)_4 \beta^3 +  \left( 5\tfrac{1}{r-1}\right)_4 \beta^{5} ~,
\end{eqnarray}
and observe that
\begin{equation}
\begin{split}
c_n 
=&  (-1)^{n} \sum_{k=0}^{n} \binom{n}{k} \left( (2k-n) \tfrac{1}{r-1}\right)_{n-1} \left[ e^{-\frac{r}{r-1}t} (2\cosh(\psi))^{\frac{1}{r-1}} \right]^{2k-n} ~.
\end{split}
\tag{\ref{4-10-2}}
\end{equation}

In the regime of interest \eqref{4-5}, $2\cosh (\psi) \approx e^{\psi}$ and therefore the series expression of $y$ (and therefore $(x_-^{+})^{r-1}$) is approximately
\begin{equation}
\begin{split}
 y = 1 + \sum_{n=1}^{\infty} \frac{(-1)^{n} e^{-n \frac{r(\psi-t)}{r-1}}}{n!}  \sum_{k=0}^{n} \binom{n}{k} \left( (2k-n) \tfrac{1}{r-1}\right)_{n-1}  e^{2k\frac{-rt+\psi }{r-1}}  ~,
\end{split}
\end{equation}
Applying Gauss's multiplication formula \eqref{a-16}, as well as Stirling's formula, we can estimate that
\begin{equation}
\begin{split}
 \frac{1}{n!} \left| \left( (2k-n) \tfrac{1}{r-1}\right)_{n-1} \right|=&  \frac{ n^{n-1} \prod_{j=2}^n \left|  \frac{2k-n}{n} \frac{1}{r-1} -1 +\frac{j}{n} \right| }{ \sqrt{2\pi} n^{n +1/2} e^{-n} } \\
< & \frac{n^{-3/2}e^n}{\sqrt{2\pi}} \prod_{j=2}^n \left( \left| \frac{2k-n}{n} \frac{1}{r-1} \right|+ \left| 1 -\frac{j}{n} \right| \right)^{n-1} \\
<& \frac{n^{-3/2}e^n}{\sqrt{2\pi}} \left( \frac{r}{r-1} \right)^{n-1} \\
\end{split}
\end{equation}
for large $n $. Thus, the absolute value of the $n$-th term of $y$ is bounded by
\begin{equation}
\begin{split}
& \frac{n^{-3/2}e^n}{\sqrt{2\pi}} \left( \frac{r}{r-1} \right)^{n-1} e^{-n \frac{r(\psi-t)}{r-1}} \left( 1 + e^{2\frac{-rt+\psi }{r-1}} \right)^n\\
 & = \frac{1}{\sqrt{2\pi}} e^{n - \frac{3}{2} \log n + (n-1) \log \frac{r}{r-1} } \left( e^{- \frac{r(\psi-t)}{r-1}} +  e^{\frac{-rt+(2-r)\psi }{r-1}} \right)^n~,
\end{split}
\end{equation}
which leads to a sufficient condition for the convergence of the series expression of $y$ (and therefore $(x_-^+)^{r-1}$ \eqref{4-10}): 
\begin{equation}
\psi -t , t - \frac{2-r}{r} \psi  >  \frac{r-1}{r} \left( 1 + \log \frac{r}{r-1} \right)~.
\label{a-36}
\end{equation}
From the fact that the R.H.S. of \eqref{a-36}, $\frac{r-1}{r} \left( 1 + \log \frac{r}{r-1} \right)$, approaches its maximum $1$ as $r \to \infty$, it can be seen that \eqref{a-36} is always satisfied if regime \eqref{4-11} is considered. That is, in the regime \eqref{4-11}, the series expression of $(x_-^{+})^{r-1}$ given in \eqref{4-10} and \eqref{4-10-2} converges.

\section{Turning point and intersections near turning point} \label{app-b}

In this appendix we figure out the position of turning point \eqref{minimalF} and show that this point satisfies \eqref{4-4}. Furthermore, near-turning-point solutions \eqref{solnnear} are calculated at the end of this section.

We start from the definition of the turning point \eqref{turnpt}, which leads to
\begin{equation}
\sinh (\theta_{\star}) +\frac{1}{2} (r-1) e^{rt}e^{(r-1)\theta_{\star} }=0 ~,
\label{b-1}
\end{equation}
where we have used the notation defined in \eqref{1-5} to rewrite the equation. It is straightforward to see that when $r=1$, $\theta_{\star} =0$. As for $r \ne 1$, we move the hyperbolic function to the R.H.S. and take the logarithm of both sides
\begin{equation}
\log \left( \sinh (\pm \theta_{\star}) \right) - (r-1) \theta_{\star}= \log \frac{| r-1 |}{2} + rt ~,
\label{b-2}
\end{equation}
where for $r <1 $ case, one takes the positive sign, and for $r>1$, the negative sign. 

Since it is the large-$t$ behaviour (or more precisely, large $|r-1| e^{rt}$ behaviour) that we are interested in, the L.H.S. of \eqref{b-2} is large and therefore \eqref{b-2} becomes a linear equation
\begin{equation}
(\pm 1 +1 -r) \theta_{\star} = \log |r-1| + rt~,
\end{equation}
of which the solution is
\begin{equation}
\theta_{\star} = \operatorname{sgn} (1- r) \frac{\log |r-1| +rt }{1+|r-1|} 
= 
\begin{cases}
\frac{\log (1-r) + rt}{2-r}~, & r<1 \\
0 ~, & r =1 \\
- \frac{\log (r-1) + rt}{r} ~, & r>1 \\ 
\end{cases}~,
\label{b-4}
\end{equation}
after taking the $r=1$ case into account. Note that \eqref{b-4} is equivalent to \eqref{minimalF}.

Furthermore, we can figure out the constant-$x$ line tangent to the turning point and prove \eqref{4-4}: For $r=1$, 
 \begin{equation}
 \begin{split}
 x(\alpha_{\star} )  =&  \frac{1}{2}\left( x_-^* + (x_-^*)^{-1}\right) + \frac{1}{2} e^{rt} (x_-^*)^{r-1} = \frac{1}{2} \left( e^{-\theta_{\star}} + e^{\theta_{\star}} \right) + \frac{1}{2} e^{t}  \\
 = & 1 + \frac{1}{2} e^{t} < \frac{1}{2} + e^{t}~.
 \end{split}
 \end{equation}
For $r \ne 1$, in general we have
\begin{equation}
\begin{split}
x(\alpha_{\star} )  =&  \frac{1}{2}\left( x_-^* + (x_-^*)^{-1}\right) + \frac{1}{2} e^{rt} (x_-^*)^{r-1} = \frac{1}{2} \left( e^{-\theta_{\star}} + e^{\theta_{\star}} \right) + \frac{1}{2} e^{rt} e^{(r-1) \theta_{\star}} \\
=& \frac{1}{2} \exp \left( \operatorname{sgn} (r-1) \frac{\log |r-1| +rt }{1+|r-1|}  \right) + \frac{1}{2} \exp \left( \operatorname{sgn} (1-r) \frac{\log |r-1| +rt }{1+|r-1|}  \right) \\
&  + \frac{1}{2}  \exp \left( rt -|r-1| \frac{\log |r-1| +rt }{1+|r-1|} \right) \\
< & \frac{1}{2} + \frac{1}{2}  \exp \left(  \frac{\log |r-1| +rt }{1+|r-1|}  \right) +  \frac{1}{2|r-1|} \exp \left( \frac{\log |r-1| +rt }{1+|r-1|} \right) ~.
\end{split}
\label{b-6}
\end{equation}
Now when $r<1$, \eqref{b-6} can be further simplified into
\begin{equation}
\begin{split}
x(\alpha_{\star} )  < & \frac{1}{2}  + \frac{1}{2} \left( 1+ \frac{1}{1-r} \right) \exp \left( \frac{\log (1-r) +rt }{2-r} \right) \\
< &  \frac{1}{2}  + \frac{1}{2} \left( 1+ \frac{1}{1-r} \right) \exp \left( \log (1-r) + rt \right)  \\
=&  \frac{1}{2} + \frac{1}{2} (2-r) e^{rt} <  \frac{1}{2} + e^t
\end{split}
\end{equation}
and when $r>1$, 
\begin{equation}
\begin{split}
x(\alpha_{\star} )  < & \frac{1}{2} + \frac{1}{2}\left( 1+ \frac{1}{r-1} \right) \exp \left( \frac{\log (r-1) +rt }{r} \right) \\
= & \frac{1}{2} +\frac{1}{2} \left( 1+ \frac{1}{r-1} \right) (r-1)^{1/r} e^t \le \frac{1}{2} + e^t~.
\end{split}
\end{equation}
since $\left( 1+ \frac{1}{r-1} \right) (r-1)^{1/r}$ reaches its maximum $2$ at $r=2$. To sum up, for all positive $r$, it is a fact that
\begin{equation}
x( \alpha_{\star} ) < \frac{1}{2} + e^t~.
\tag{\ref{4-4}}
\end{equation}

We close this section by determining the intersections of the Fermi surface and a constant-$x$ line that satisfies
\begin{equation}
x = x(\alpha_{\star}) + \delta x, \qquad \frac{\delta x}{ x(\alpha_{\star})} \ll 1~.
\end{equation}
In this case, \eqref{1-3} becomes
\begin{equation}
\frac{1}{x_-} + x_- + e^{rt} x_-^{r-1} = \frac{1}{x_-^*} + x_-^* + e^{rt} (x_-^*)^{r-1}  + 2\delta x~.
\label{b-10}
\end{equation}
Expand $x_-$ around $x_-^*$ as
\begin{equation}
x_- = x_-^* \left( 1 + u (\delta x)^{1/2} +\mathcal{O} (\delta x) \right)~,
\label{b-11}
\end{equation}
substitute it into \eqref{b-10}, and we obtain a quadratic equation of $u$ 
\begin{equation}
\frac{1}{x_-^*} u^2 \delta x +\frac{(r-1)(r-2)}{2} e^{rt} (x_-^*)^{r-1}u^2 \delta x = 2 \delta x~.
\end{equation}
Therefore, the two solutions of \eqref{1-3} near turning point are
\begin{equation}
x_-^{\pm} = x_-^* \left[ 1 \pm \left( \frac{2\delta x}{ (x_-^*)^{-1}  +\frac{(r-1)(r-2)}{2} e^{rt} (x_-^*)^{r-1} } \right) + \mathcal{O} (\delta x) \right]~.
\tag{\ref{solnnear}}
\end{equation}

\section{Cubic couplings at the boundary}\label{app-c}

In this appendix we provide calculational details of the evaluation of the cubic couplings in section \ref{divcp}. Enlightened by appendix \ref{app-a}, we can further divide the regime of interest \eqref{4-5} into \eqref{4-8} and \eqref{4-11} and consider the cubic couplings of any positive $r$ in these two regimes respectively. 

Below we use coordinates $(t,q)$ instead of $(t, \psi)$, in terms of which the two regimes, \eqref{4-8} and \eqref{4-11}, become
\begin{equation}
\begin{cases}
q \gg t \gg 1 ~, & r<1\\
\frac{2-r}{r} q \gg t \gg 1 ~, & 1<r<2\\
\end{cases}, \qquad \text{i.e.}\quad 
\frac{1-|r-1|}{r} q \gg t \gg 1, \quad 0< r <2
\label{c-1}
\end{equation}
and
\begin{equation}
\begin{cases}
t \gg \frac{2-r}{r} q \gg 1, & 1<r<2 \\
t, q \gg 1, &  r \ge 2 \\
\end{cases}, \qquad \text{i.e.} \quad q \gg 1 ~\& ~ t \gg \max \left\{ 1, \frac{2-r}{r} q \right\}, \quad r >1~,
\label{c-2}
\end{equation}
respectively, according to section \ref{minkowskiancoord}.

\subsection{In the regime $\frac{1-|r-1|}{r} q \gg t \gg 1$ where $0< r <2$}

In this regime we can summarise from \eqref{5-1} and the first line of \eqref{5-2} (or equivalently, the first line of \eqref{5-3}) that the Minkowskian coordinates are 
\begin{equation}
q = \psi ,\qquad \tau =t~.
\end{equation}
Hence the coordinates $(t,q)$ are related to $P_{\pm}$ via \eqref{4-20} and the first line of \eqref{w-soln}, in particular,
\begin{equation}
P_{\pm}(q,t) = \pm \sinh (q) + \frac{1}{2} \exp \left( rt \mp (r-1) q \right)~.
\end{equation}
Utilising \eqref{1-8a}, the collective field and the conjugate momenta are therefore
\begin{eqnarray}
\partial_x \varphi_0 &=& \frac{1}{2\pi} (P_+ - P_-) =\frac{1}{2\pi}\left(  2\sinh (q) - e^{rt} \sinh ( (r-1) q) \right)~, \label{c-5}\\
\frac{\partial_t \varphi_0}{\partial_x \varphi_0} &=& -\frac{1}{2} (P_+ + P_-) =  -\frac{1}{2} e^{rt}\cosh ( (r-1) q)~.\label{c-6}
\end{eqnarray}
We can simplify \eqref{c-5} by keeping only the dominant term $\sinh (q)$, of which the dominance is recognised by
\begin{equation}
\frac{e^{rt} \sinh ( (r-1) q) }{\sinh (q) }  \sim e^{rt + |r-1|q -q}  = \exp \left[ r \left( t - \frac{1-|r-1|}{r} q \right) \right] \ll 1
\label{ratio}
\end{equation}
according to \eqref{c-1}.
Noting that $e^{rt} \sinh ( (r-1) q)  $ and $e^{rt}\cosh ( (r-1) q)$ are at the same order when $q \gg 1$, i.e.
\begin{equation}
e^{rt}\cosh ( (r-1) q) \sim e^{rt} \sinh ( (r-1) q)   \sim e^{rt + |r-1| q}~,
\end{equation}
from \eqref{ratio} we also learn that
\begin{equation}
\frac{\partial_t \varphi_0}{(\partial_x \varphi_0)^2} \sim \frac{e^{rt} \cosh ( (r-1) q) }{\sinh (q) }  \sim e^{rt + |r-1|q -q}   \ll 1~.
\end{equation}
Thus, it can be seen that all the three cubic couplings are small, since
\begin{equation}
\frac{1}{(\partial_x \varphi_0)^2} \sim e^{-2q},\qquad 
\frac{\partial_t \varphi_0}{(\partial_x \varphi_0)^3} = \frac{1}{\partial_x \varphi_0} \cdot \frac{\partial_t \varphi_0}{(\partial_x \varphi_0)^2} \sim e^{rt + |r-1|q -2q}, \qquad 
 \frac{(\partial_t \varphi_0)^2}{(\partial_x \varphi_0)^4} \sim  e^{2(rt + |r-1|q -q)} 
  \label{cubic-2}
\end{equation}
\eqref{cubic-2} is equivalent to \eqref{7-1} and \eqref{7-2} when $0<r<1$ and $1<r<2$ respectively.

\subsection{In the regime $q \gg 1$ and $t \gg \max \left\{ 1, \frac{2-r}{r} q \right\}$ where $r \ge 1$}

In this regime we can summarise from the second line of \eqref{5-2} (or equivalently, the second line of \eqref{5-3}) and \eqref{5-4}  that the Minkowskian coordinates are 
\begin{equation}
q = \frac{1}{2} \frac{r}{r-1} (\psi - t), \qquad \tau = \frac{1}{2} \frac{r-2}{r-1} ( t- \psi)~.
\end{equation}
The coordinates $(t,q)$ are related to $P_{\pm}$ via the second line of \eqref{w-soln} and \eqref{w-soln2}, in particular,
\begin{eqnarray}
P_+(q,t) 
&=&  \sinh \left(t + \frac{2(r-1)}{r} q \right) +\frac{1}{2} \exp \left( t  - \frac{2(r-1)^2}{r} q \right)~, \\
P_-(q,t) 
&=&  \sinh \left( t - \frac{2}{r}q \right) + \frac{1}{2}\exp \left( t  + \frac{2(r-1)}{r} q \right) ~.
\end{eqnarray}
The collective field and the conjugate momenta are therefore
\begin{eqnarray}
\partial_x \varphi_0 &=& \frac{1}{2\pi} \left[ - \frac{1}{2} e^{ - t  - \frac{2(r-1)}{r} q } +\frac{1}{2} e^{ t  - \frac{2(r-1)^2}{r} q } -  \sinh \left( t - \frac{2}{r}q \right)  \right] ~,\\
\frac{\partial_t \varphi_0}{\partial_x \varphi_0} &=& -\frac{1}{2} \left[  e^{  t + \frac{2(r-1)}{r} q } -  \frac{1}{2} e^{ - t  - \frac{2(r-1)}{r} q } +\frac{1}{2} e^{ t  - \frac{2(r-1)^2}{r} q } +  \sinh \left( t - \frac{2}{r}q \right) \right] ~.
\end{eqnarray}

Similar to the previous section, we only keep the dominant terms determined as follows: 
For $\frac{\partial_t \varphi_0}{\partial_x \varphi_0}$, it is not difficult to see that 
\begin{equation}
\frac{\partial_t \varphi_0}{\partial_x \varphi_0} = -\frac{1}{2}  e^{  t + \frac{2(r-1)}{r} q } ~,
\end{equation}
since
\begin{eqnarray}
\frac{e^{ - t  - \frac{2(r-1)}{r} q}}{e^{  t + \frac{2(r-1)}{r} q }} \sim e^{-2 t } e^{ 2 \frac{2(r-1)}{r}q} \ll 1, &\quad& 
\frac{e^{ t  - \frac{2(r-1)^2}{r} q }}{e^{  t + \frac{2(r-1)}{r} q }} \sim e^{-2 (r-1)q} \ll 1, \nonumber \\
\frac{e^{ t  - \frac{2}{r} q }}{e^{  t + \frac{2(r-1)}{r} q }}  \sim e^{-2q} \ll 1, & \quad & 
\frac{e^{ -t  + \frac{2}{r} q }}{e^{  t + \frac{2(r-1)}{r} q }} \sim e^{-2 \left( t -\frac{2-r}{r}q \right) } \ll 1,
\label{c-17}
\end{eqnarray}
according to \eqref{c-2}. Finding the dominant terms of $\partial_x \varphi_0$ requires more effort, but actually we have already learned from \eqref{c-17} that
\begin{equation}
 \frac{\partial_t \varphi_0}{(\partial_x \varphi_0)^2} \sim \frac{ e^{t + \frac{2(r-1)}{r}q } }{  \frac{1}{2} e^{ - t  - \frac{2(r-1)}{r} q } - \frac{1}{2} e^{ t  - \frac{2(r-1)^2}{r} q } +  \sinh \left( t - \frac{2}{r}q \right)  } \gg 1~.
\end{equation}
This implies that there exists at least one divergent cubic coupling $ \frac{(\partial_t \varphi_0)^2}{(\partial_x \varphi_0)^4} $.

Now for completeness, we figure out the dominant terms of $\partial_x \varphi_0$ for $1<r<2$ and $r>2$, separately. We first argue that $e^{ - t  - \frac{2(r-1)}{r} q }$ is a subleading term for all $r>1$, which can been see from
\begin{equation}
\frac{ e^{ - t  - \frac{2(r-1)}{r} q }}{ e^{-t + \frac{2}{r}q} } \sim e^{-2 q} \ll 1~.
\end{equation}
Moreover, when $1<r<2$, $e^{t - \frac{2}{r}q}$ is also small, since
\begin{equation}
\frac{ e^{t - \frac{2}{r}q} }{ e^{ t  - \frac{2(r-1)^2}{r} q } } \sim e^{ 2 (r-2) q } \ll 1~,
\end{equation}
which also tells us that when $r>2$, $e^{ t  - \frac{2(r-1)^2}{r} q }$ is instead the small one. The two terms are at the same order when $r=2$ and they cancel each other in $\partial_x \varphi_0$. To sum up, $\partial_x \varphi_0$ can be simplified into
\begin{equation}
\partial_x \varphi_0 \approx \frac{1}{2\pi} 
\begin{cases}
\frac{1}{2} e^{ t  - \frac{2(r-1)^2}{r} q } + \frac{1}{2} e^{- t + \frac{2}{r}q } ~,& 1<r<2 \\
\frac{1}{2} e^{ -t + q } ~,& r=2 \\
   \sinh \left( -t + \frac{2}{r}q \right) ~,& r>2 \\
\end{cases}~.
\end{equation}
The resulting cubic couplings are given in \eqref{7-3}, \eqref{7-4} for $1<r<2$ and in \eqref{7-5}, \eqref{7-6} for $r>2$.

\end{document}